\RequirePackage{fix-cm}
\documentclass[smallextended]{svjour3}       
\smartqed  
\usepackage{graphicx}
\usepackage{cite}

\usepackage{ocgx2}

\begin{document}

\title{Spectral engineering via complex patterns of circular nano-object miniarrays: II. concave patterns tunable by integrated lithography realized by circularly polarized light
}

\titlerunning{Spectral engineering via complex concave patterns}        

\author{Emese Toth          \and
        Aron Sipos			\and
        Oliver A. Fekete 	\and
        Maria Csete* 
}


\institute{Emese Toth \at
              Department of Optics and Quantum Electronics, \\
              University of Szeged, H-6720, Szeged Dóm tér 9, Hungary \\
		   \and
		   Aron Sipos \at
			  Institute of Biophysics, Biological Research Centre, \\
			  H-6726 Szeged Temesvári krt. 62, Hungary \\
              \email{sipos.aron@brc.hu}           
           \and
           Oliver A. Fekete \at
              Department of Optics and Quantum Electronics, \\
              University of Szeged, H-6720, Szeged Dóm tér 9, Hungary \\
           \and
           *Maria Csete \at
              Department of Optics and Quantum Electronics, \\
              University of Szeged, H-6720, Szeged Dóm tér 9, Hungary \\
              \email{mcsete@physx.u-szeged.hu}           
}


\maketitle

\begin{abstract}

Application of circularly polarized beams in interferometric illumination of colloid sphere monolayers enables the direct fabrication of rectangular patterns consisting of circular nanohole miniarrays in metal films. The spectral and near-field effects of complex rectangular patterns made of a central nanoring and slightly rotated satellite nanocrescents were studied in azimuthal orientations promoting the localized and propagating plasmon’s coupling. To inspect the localized modes separately, spectral responses and near-field phenomena of hexagonal patterns composed of uniform nanorings and nanocrescents, that can be fabricated by a perpendicularly and obliquely incident single homogeneous circularly polarized beam, were investigated. To uncover the interaction of localized and propagating modes, artificial rectangular patterns composed of a singlet nanoring, singlet horizontal nanocrescent and quadrumer of four slightly rotated nanocrescents were analyzed. It was demonstrated that the interacting C2 and C1 localized resonances on the (approximately) horizontal nanocrescents in C orientation (($16^{\circ}$) $0^{\circ}$ azimuthal angle) and the azimuthal orientation (in)dependent localized resonance on the (nanorings) nanocrescents coupled with propagating surface plasmon polaritons (close to) in U orientation (($106^{\circ}$)$90^{\circ}$ azimuthal angle) result in similar split spectra. The spectral response of the complex miniarray pattern can be precisely tuned by varying the geometrical parameters of the moderately interacting nanoholes and the pattern period. Enhancement of dipolar emitter’s fluorescence is demonstrated in appropriate configurations that have a potential application in bio-object detection.
\end{abstract}

\section{Introduction}
\label{intro}

The surface plasmons were rediscovered due to the recognition of unique spectral and near-field properties of hole-arrays in metal films \cite{ref_01,ref_02,ref_03,ref_04,ref_05}. Such concave patterns can act as plasmonic crystals, and the resulted optical properties can be tuned by controlling the shape and size parameters either of the composing nanoholes or of the periodic pattern.\\
An individual nanohole inside a metal film can already result in a spectral modulation and a significant $\bar{E}$-field enhancement depending on it’s size, shape and relative orientation with respect to the $\bar{E}$-field oscillation direction \cite{ref_02,ref_04,ref_05}. Among different geometries C-shaped apertures are particularly interesting, since they can result three orders of magnitude $\bar{E}$-field enhancement and a confinement down to the tenth of the wavelength, which was first demonstrated in the microwave region \cite{ref_04,ref_06}. \\
On the arrays of sub-wavelength holes with different shapes extraordinary transmission can occur \cite{ref_01,ref_02,ref_03,ref_04,ref_05}. The earliest studies revealed that transmittance minimum / maximum appears at the spectral position corresponding to the Rayleigh / resonant Wood anomaly, the former / latter is related to the photonic mode scattering at grazing angle / propagating surface plasmon polariton (SPP) excitation \cite{ref_07,ref_08}. Important difference with respect to the photonic crystals is that the SPPs propagate in the plane of topographic modulation on hole-arrays acting as plasmonic cystals, which are embedded into lossy media. It was proven that the presence of nanoholes causes band-bending and appearance of band-gaps, indicating that application of a simple planar interface approximation to explain the dispersion branches can cause discrepancies \cite{ref_09,ref_10}. At the band-gaps standing SPP waves were detected on rectangular arrays of sub-wavelength holes via SNOM \cite{ref_11}. \\
The main design rules of plasmonic spectral engineering were uncovered, which rely on the interplay between the Fabry-Perot resonances localized inside individual nanoholes, resulting in broad spectral lines, and grating-coupled resonances on plasmonic lattices, resulting in narrow spectral features \cite{ref_03,ref_04,ref_05}. In two-dimensional hole-arrays the periodicity along a certain direction is at play, when the $\bar{E}$-field projection is significant along the corresponding $\bar{k}$ lattice vector, since the parallelism promotes the SPP-grating coupling phenomenon. Narrow spectral features originating from higher order Bragg resonances were identified on microscale rectangular arrays of nanoholes \cite{ref_12}. It was demonstrated that the coupled LSPR and SPP can result in a complex and finely tunable spectral response \cite{ref_13}. \\
The varieties of concave patterns in metal films have already different important applications. The coupled resonance resulted narrow Fano-lines are particularly beneficial in bio-sensing applications \cite{ref_14}. The polarization insensitivity in square arrays of spherical nanoholes makes it possible to preserve polarization entanglement \cite{ref_15}. Moreover, elliptical holes arranged in two sub-lattices enable polarization induced frequency shifts as well \cite{ref_16}. \\
The first metamaterials, which act close to the visible region, were created by using rectangular arrays of sub-wavelength holes \cite{ref_17}. C-shaped apertures were applied to enhance the photocurrent in Ge detectors at 1310 nm, and maximal efficiency was experienced in case of $\bar{E}$-field oscillation direction parallel to the arms \cite{ref_18}. Several examples prove that the widely tunable resonances on nanovoid and nanohole-arrays enable uniquely high sensitivity bio-sensing \cite{ref_19}. Combined spectral and near-field studies revealed that the transmittance minima (maxima) exhibit smaller (larger) sensitivity on hexagonal as well as on square hole-arrays \cite{ref_20,ref_21,ref_22}. Moreover, special individual scatterers, such as hole-doublets ensure enhanced sensitivity due to the antennas appearing at their apexes \cite{ref_04,ref_23}. Transmittance peaks corresponding to Bragg resonance along the $\bar{E}$-field oscillation direction exhibit considerably enhanced sensitivity on square arrays \cite{ref_24}. The high local field-enhancement achievable via nanorings embedded into continuous metal film has a potential application in SERS \cite{ref_25}. Different types of nano-apertures in metal films enhance fluorescence efficiency, which promotes single molecule detection \cite{ref_26,ref_27}. The common advantage of structures with a rectangular unit cell is that typically smaller size enables them to be effective in the NIR region, which is especially important in bio-detection \cite{ref_28}. Strong-coupling regime of photochromic molecules can be reached on a hole-array platform \cite{ref_29}. Steady-state superradiance can be achieved by coupling the SPPs on a two-dimensional hole-array via dye molecules arranged above them \cite{ref_30}. Moreover, lasing of a semiconductor gain medium governed by metal hole-array dispersion was demonstrated as well \cite{ref_31,ref_32}. \\ 
Complementary C-shaped split ring resonators were proposed to design metasurfaces with potential applications as frequency and polarization selective filters \cite{ref_33}. Directional coupling of SPP waves was performed by arrays of complementary C-shaped split-ring-resonators, which is advantageous for miniaturizing photonic and plasmonic circuits \cite{ref_34}. Babinet inverted plasmonic metasurfaces were used to produce spin-selective second harmonics vortex beams \cite{ref_35}. It was shown that disks and cylindrical apertures show quantitative differences with respect to the Babinet complementarity, since the apertures exhibit higher magnetic-to-electric field ratio, and better heat and charge transfer properties \cite{ref_36}. Miniarrays of elliptical nanoholes that are not superimposable on their mirror image, exhibit chiroptical effects, e.g. selectively reflects one spin state, while preserving its handedness \cite{ref_37}.\\
However, fabrication of non-hexagonal, e.g. rectangular arrays of nano-objects was previously possible only by the expensive and complex e-beam lithography procedures \cite{ref_38}. Double exposure two-beam interference lithography has been applied to fabricate 1D and 2D structures in gold film, e.g. rectangular array of nanoholes \cite{ref_39}. Colloid sphere lithography has been used to generate imprinted nanochannel alumina (NINA) pattern with a hexagonal symmetry \cite{ref_40}.  The laser based colloid sphere lithography is capable of fabricating nanoholes via colloid spheres, which can be isolated, aggregated and embedded into monolayers, and the fabricated nanohole parameters depend on the colloid sphere’s size, material, on the environment and substrate as well as on the laser parameters \cite{ref_41}. The tilting during treatment makes it possible to fabricate sub-diffraction objects of arbitrary shape \cite{ref_42}. As an alternative method, combination of phase-shifting lithography, etching, e-beam deposition and lift-off of the film (PEEL) was developed, which was used to fabricate microscale periodic rectangular arrays of nanoscale holes \cite{ref_12}. Nanorings have been also prepared by bio-molecule assisted deposition of gold spheres into previously fabricated nanoholes \cite{ref_25}. \\
To overcome the limits of laser based colloid sphere lithography, chemical treatments were applied to create Janus and patchy colloids, template based substrates were fabricated to ensure geometrical confinement, and electric as well as magnetic forces were applied to control location of colloid assemblies \cite{ref_43}. Multiscale periodic colloidal assemblies were created by a method that combines thermo-responsive depletion with pre-patterned surface features \cite{ref_44}. \\
In our previous studies we have presented the interferometric illumination of colloid sphere monolayers (IICSM) that enables to tune large number of geometrical parameters independently \cite{ref_45,ref_46,ref_47,ref_48}. The method combines colloid sphere (CS) and interference lithography (IL) and synthesizes all advantages of these methods, thus providing a good tool to fabricate versatile patterns. The IICSM method has been already presented for illumination by linearly \cite{ref_45,ref_46} and circularly \cite{ref_47,ref_48} polarized light in our previous papers. \\
In this paper we present the spectral and near-field effects achievable by complex patterns of rounded nanoholes, namely by miniarrays of nanorings and nanocrescents that form an ordered and pre-designed rectangular pattern. The rounded objects originate from illumination by circularly polarized beams and form miniarrays defined by the specific interferometric illumination configuration. For reference purposes spectral and near-field effects achievable by hexagonal patterns of nanorings and nanocrescents, that originate from illumination of colloid sphere monolayers by a single perpendicularly and obliquely incident homogeneous circularly polarized beam, are also presented. The spectral and near-field effects of the building blocks in rectangular patterns, namely a singlet nanoring, a singlet horizontal nanocrescent and a quadrumer of slightly rotated satellite nanocrescents, are also analyzed. The potential of all these structures to enhance fluorescence of dipolar emitters is also demonstrated. Part I of this paper describes the spectral and near-field effects of the analogue convex patterns \cite{ref_49}. A comparative study on the spectral and near-field effects of complementary concave and convex patterns, that can be directly fabricated via IICSM and by a following lift-off procedure, has already been published in a former conference paper \cite{ref_48}. A detailed comparative study on the corresponding optical responses of complementary patterns, namely on the reflectance (transmittance) of the concave and on the transmittance (reflectance) of the convex patterns, is presented in our upcoming paper \cite{ref_50}.

\section{Method}
\label{Met}


\subsection{Numerical modeling and characterization of patterns consisting of concave spherical nano-objects}
\label{Num_mod}

\begin{figure}[h]
\center
	{\includegraphics[width=0.75\textwidth]{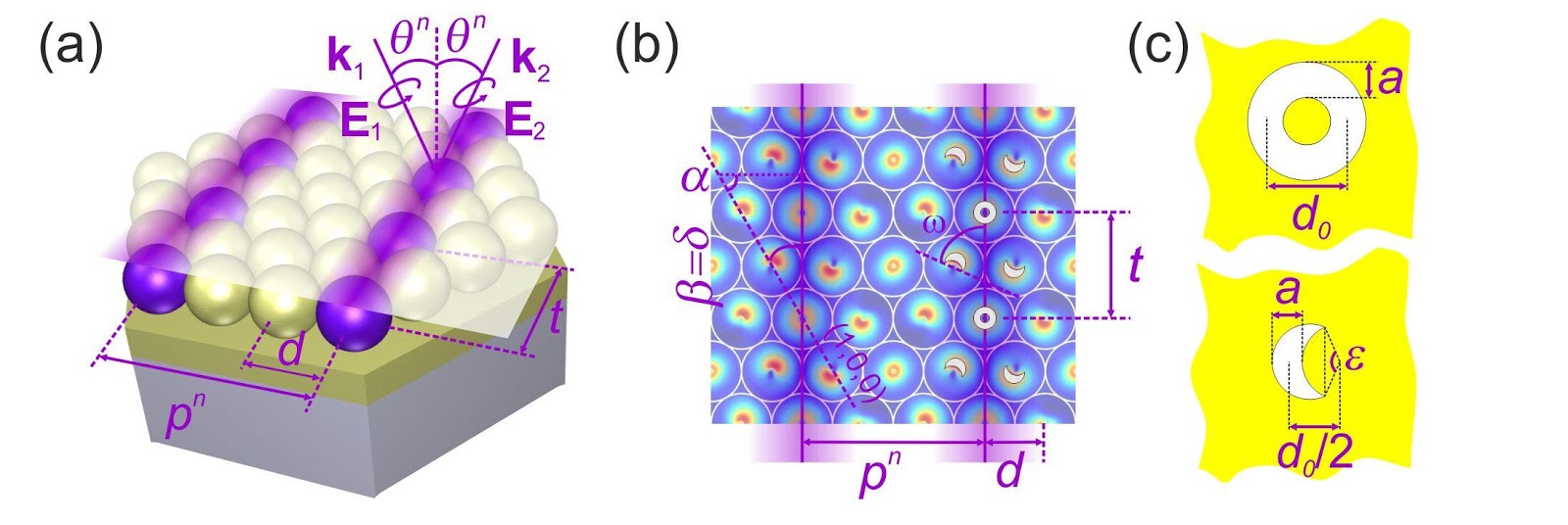}}
\caption{Schematic drawing and characteristic size parameters (a) 3D scheme of $p^n$ pattern parameter variation (b) by modifying $\theta^n$ polar angle orientation of incidence plane ($\alpha$), and the resulted interference pattern ($\beta$) to tune $d$ colloid sphere diameter-scaled $t$ distance of nano-objects, (c) generated features qualified by $d_0$, $a$ size $\varepsilon$ shape and $\omega$ orientation parameters.}
\label{colloid_fig_01}       
\end{figure}

We used finite element method (FEM), namely the Radio Frequency module of COMSOL Multiphysics software package (COMSOL AB, Sweden) to study the structures that can be fabricated by a homogeneous beam illumination and by IICSM using circularly polarized light. In this study $d’$=100 nm diameter Au colloid spheres are arranged into a hexagonally close-packed monolayer on a 45 nm thick gold film covered NBK7 substrate, and are illuminated by a single perpendicularly or obliquely incident, or by two obliquely incident interfering $\lambda$=400 nm circularly polarized beams. The illumination of one hexagonal Au colloid sphere monolayer unit cell was performed by using 3D periodic models, which enable the setting of any desired azimuthal orientation and angle of incidence for a single and as well as for multiple beams. All materials were modeled by taking into account wavelength dependent optical properties, namely the NBK7 glass was qualified by the Cauchy formulae $n=A+\frac{B}{\lambda^2}+\frac{C}{\lambda^4}$ ($A_{NBK7}$= $1.503$, $B_{NBK7}$ = $4.395\cdot10^{-3}$, $C_{NBK7}$ = $-8.585\cdot10^{-5}$) \cite{ref_51}, whereas for gold tabulated data sets were interpolated to implement the wavelength dependent dielectric properties \cite{ref_52}.\\
The schematic drawing in Fig. \ref{colloid_fig_01}a shows the IICSM concept for the case of illumination by two circularly polarized beams. The required condition of IICSM method realization is the perfect synchronization of a hexagonally closely packed colloid sphere monolayer and an illuminating interference pattern. Fig. \ref{colloid_fig_01}b  and c show the main characteristic geometrical parameters that are variable via IICSM, when it is realized by circularly polarized beam. In so-called closely packed in between arrays configuration the $p^n$ periodicity can be tuned to $p^n=n\cdot d/2$ discrete values, which are achievable at the incidence angle of $\theta^n=arcsin(\lambda/(n\cdot d))$, where $d$ is the colloid spheres diameter, and $n\ge1$ is an integer. The $\alpha$ angle shows the incidence plane orientation with respect to the (100) lattice direction of the hexagonal monolayer, while the resulted $t$ inter-object distance varies with the corresponding $\beta$ orientation of the interference pattern. Fig. \ref{colloid_fig_01}b and c show the complex interference pattern with $t=\sqrt{3}\cdot d$ inter-object distance that can be fabricated in case of $\beta$=3$0^{\circ}$, and the resulted concave pattern in an Au layer, which is inspected in this work. The $d_0$ and $a$ nano-object size parameters are tunable by the $\lambda$ wavelength and beam power density, as well as by the $d'$ colloid sphere diameter and its material. In our present study $d$=$d’$=100 nm, which implies that we suppose illumination of touching gold colloid spheres. This can be supposed, since in our previous studies it has been proven that the nano-object parameters do not modify significantly, when the diameter is reduced to ensure larger intensity at the substrate interface than at the monolayer central plane, where the spheres touch each other. Both the nanoring and nanocrescent shaped nanoholes stem from the circular polarization of the beams, proving that the exact shape of generated nano-objects is polarization dependent, according to our previous studies \cite{ref_40,ref_41,ref_42,ref_43,ref_44,ref_45,ref_46,ref_47,ref_48}.

\subsection{Spectral and near-field study of different patterns}
\label{Spe_nea_fie_stu}
The complete spectral study of 45 nm thick gold films decorated by different nanohole patterns was performed, i.e. the effect of these complex patterns on the optical response and on the near-field distribution was determined. Floquet boundary condition was applied on the vertical boundaries of FEM (COMSOL) models consisting of different hexagonal and rectangular unit cells during p-polarized plane wave illumination. The inspected realistic hexagonal patterns are as follows: hexagonal pattern of (i) nanoring shaped holes (Fig. \ref{IICSM_DG_cv_01}a, b, d), (ii) horizontal nanocrescent shaped holes (Fig. \ref{IICSM_DG_cv_01}a, c, e) (vertical nanocrescents are presented in a Supplementary material). The nanoholes are nominated as nanorings and nanocrescents for the shake of simplicity. The studied artificial composing rectangular patterns are as follows: 300 nm rectangular pattern of (iii) a singlet nanoring (Fig. \ref{IICSM_DG_cv_02}), (iv) a singlet horizontal nanocrescent (Fig. \ref{IICSM_DG_cv_03}), (v) a quadrumer of slightly rotated nanocrescents (Fig. \ref{IICSM_DG_cv_04}). Finally, two different rectangular patterns were analyzed: (vi) 300 nm (Fig. \ref{IICSM_DG_cv_05}) and (vii) 600 nm (Fig. \ref{IICSM_DG_cv_06}) rectangular pattern of the same miniarray composed of a central nanoring and a quadrumer of slightly rotated nanocrescents. \\
In this spectral study p-polarized light illuminated the perforated gold film in a symmetric environment, meaning that the film is surrounded by and the nanoholes are filled with NBK7 glass material. The hexagonal pattern of concave nanorings and nanocrescents have been inspected in $0^{\circ}$ and $90^{\circ}$ azimuthal orientations in order to uncover the characteristic LSPRs supported by the nanoholes without grating-coupling either of photonic or plasmonic modes. In case of rectangular patterns both the $90^{\circ}$/$106^{\circ}$ and $0^{\circ}$/$16^{\circ}$ azimuthal orientations have been inspected, since these promote LSPR as well as grating-coupling effects in case of horizontal nanocrescents / LSPR in case of the quadrumer and miniarrays. The LSPR on nanocrescent shaped holes is nominated as C and U resonance in case of $\bar{E}$-field oscillation direction parallel and perpendicularly to their symmetry axis, by following the nomenclature introduced in case of convex nanocrescents, but taking into account the complementarity according to the Babinet principle \cite{ref_33,ref_35,ref_36,ref_53}. Further details are provided in our corresponding papers \cite{ref_48,ref_49,ref_50}. The spectra were taken throughout the 200 nm–1000 nm interval with 10 nm resolution, at $\varphi$= $0^{\circ}$ polar angle, namely at perpendicular incidence (Fig. \ref{IICSM_DG_cv_01}-\ref{IICSM_DG_cv_06}/a).\\
FEM (COMSOL) was applied to inspect the dispersion characteristics of the concave patterns by selecting fractions on the high symmetry path throughout their IBZ according to the azimuthal orientations that promote uncovering of the LSPR on composing circular nano-objects and mapping of the coupled SPP branches. Accordingly, the dispersion characteristics have been taken in $0^{\circ}$ and $90^{\circ}$ azimuthal orientations. In case of dispersion diagram computations, the spectral range was extended through 1000 nm applying the same 10 nm wavelength resolution, as in case of perpendicular incidence, whereas the $\varphi$ incidence angle was also modified from $0^{\circ}$ to $85^{\circ}$ with $5^{\circ}$ steps (Fig. \ref{IICSM_DG_cv_01} d, e and Fig. \ref{IICSM_DG_cv_02}-\ref{IICSM_DG_cv_06}/c). Wherever needed to uncover all underlying modes on the dispersion graphs, higher resolution complementary calculations were performed with smaller steps.\\
According to the literature, in case of plasmonic patterns the absorptance spectra are the most informative to find resonances, therefore the absorptance spectra and dispersion characteristics taken in absorptance are analyzed throughout this paper \cite{ref_54}. To separate the effect of complex plasmonic structures from the background of the continuous gold film, the spectra were rectified. Namely, first the optical signal of a solid gold film having the same thickness as the perforated film was subtracted, then the absorptance values were normalized by multiplying the spectra with the (unit cell) / (nanohole) surface area ratio (Fig. \ref{IICSM_DG_cv_01}-\ref{IICSM_DG_cv_06}/a). \\
The near-field and charge distribution have been inspected by taking into account the complementarity of the modal profiles according to the Babinet principle \cite{ref_33,ref_35,ref_36}. Namely, the complementarity of the time-averaged $E_z$ distribution on convex patterns corresponding with time-averaged $B_z$ distribution on concave patterns allowed to determine the accompanying time-averaged $E_z$ distribution and to uncover the characteristic charge distribution at C and U resonances on patterns consisting of concave nanocrescents (Fig. \ref{IICSM_DG_cv_01}b, c and Fig. \ref{IICSM_DG_cv_02}-\ref{IICSM_DG_cv_06}/b) \cite{ref_48,ref_49,ref_50,ref_53}.\\ 
Localised plasmon resonances on the nanorings and nanocrecents are distinguished by using “r” and “c” in the abbreviations. SPPs grating-coupled in first and second order are distinguished by “SPP1” and “SPP2” abbreviations.\\
FEM was used to inspect the capabilities of the artificial composing patterns, namely the rectangular patterns of singlet nanoring and quadrumer of nanocrescents, as well as of the rectangular pattern of their miniarray to enhance the fluorescence (Fig. \ref{dip_encha_01}).

\section{Results and discussion}
\label{Res_dis}

\subsection{Patterns achievable in different illumination configurations}
\label{Pat_in_dif_ill}

\begin{figure}[h]
\center
	{\includegraphics[width=0.75\textwidth]{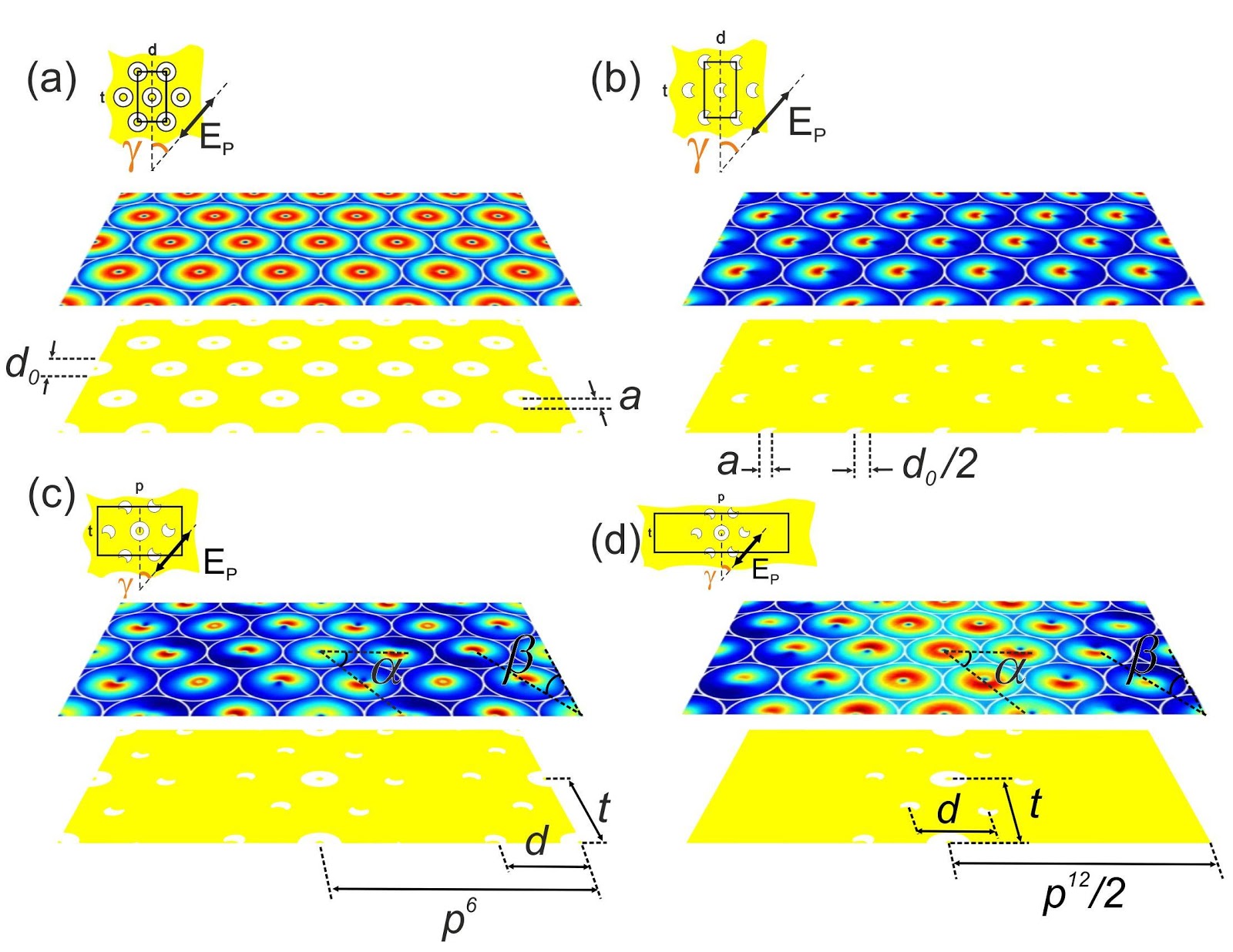}}
\caption{Normalized $\bar{E}$-field distribution under Au colloid sphere monolayers illuminated by single circularly polarized beam incident (a) perpendicularly and (b) obliquely, which results in a hexagonal pattern of nanoholes, and (c, d) by two interfering beams (in $\beta$ =$30^{\circ}$ orientation with respect to the (100) of the original colloid sphere monolayer) in IICSM configuration, which results in rectangular pattern of miniarrays. The insets show the unit cells of the complex patterns: hexagonal pattern of (a) nanorings and (b) nanocrescents, and rectangular pattern of miniarrays made of a central nanoring and satellite nanocrescents with (c) $p^6$= 300 nm and (d) $p^{12} $= 600 nm periodicity.}
\label{colloid_fig_02}
\end{figure}

To determine the pattern types achievable by homogeneous illumination and by IICSM, the $\bar{E}$-field distribution was examined at the surface of gold films. The $d_0$ and $a$ parameters were calculated based on the FWHM of the intensity distribution. Illumination by a single homogeneous, perpendicularly incident circularly polarized beam results in a hexagonal pattern of uniform nanorings with 10 nm and 46 nm inner and outer diameters, respectively (Fig. \ref{colloid_fig_02}a). Obliquely incident single homogeneous circularly polarized beam generates hexagonal pattern of uniform nanocrescents, Fig. \ref{colloid_fig_02}b indicates the case of $\theta^6= 41.8^{\circ}$ incidence angle. The nanocrescents were approximated as the intersections of two cylindrical objects with 25 nm and 20 nm diameter, with 12.5 nm center-distance. Fig. \ref{colloid_fig_02}c and $d$ illustrates the effect of interference pattern periodicity modification in IICSM. Namely, (vi) a rectangular pattern with $p^6$ = 300 nm periodicity is achievable at $\theta^6= 41.8^{\circ}$ angle of incidence corresponding to $n$ = 6 case, whereas (vii) $p^{12}$=600 nm periodic rectangular pattern develops at $\theta^{12}= 19.5^{\circ}$ incidence angle corresponding to $n$ = 12 case. Taking into account that the $d_0$ and $a$ nano-object size parameters are tunable with the power density and to simplify the comparison between miniarrays with different periodicities, analogous nanohole parameters were supposed during the spectral and near-field study of both rectangular patterns, as shown in the insets of Fig. \ref{colloid_fig_02}. Namely, the inner and outer central nanorings have similar 10 nm and 50 nm diameters, while the satellite nanocrescents have the same size parameters, as in their hexagonal array (Fig. \ref{colloid_fig_02}c, d).

\subsection{Spectral and near-field effects of different patterns}
\label{Spec_nea_fie_eff}

The common property of the rectified absorptance extracted from concave patterns is that the fingerprints of the spectrally overlapping and interacting particle plasmon resonance (PPR) and c-C2 resonance in C orientation, as well as the PPR signatures in U orientation are missing caused by the subtraction of the continuous film absorptance, which exhibits commensurate enhancement in the spectral interval of PPR \cite{ref_49,ref_50}. According to the Babinet principle it was proven that the $E_z$ field component distribution on the complementary convex pattern corresponds with the $B_z$ field component distribution on the concave pattern, and the accompanying $E_z$ field component distribution on the concave pattern revealed the charge distribution at the characteristic resonances \cite{ref_33,ref_35,ref_36,ref_49,ref_50}. 

\subsubsection{Hexagonal pattern of concave nanorings}
\label{Hex_rin}

\begin{figure}[h]
\center
	{\includegraphics[width=0.75\textwidth]{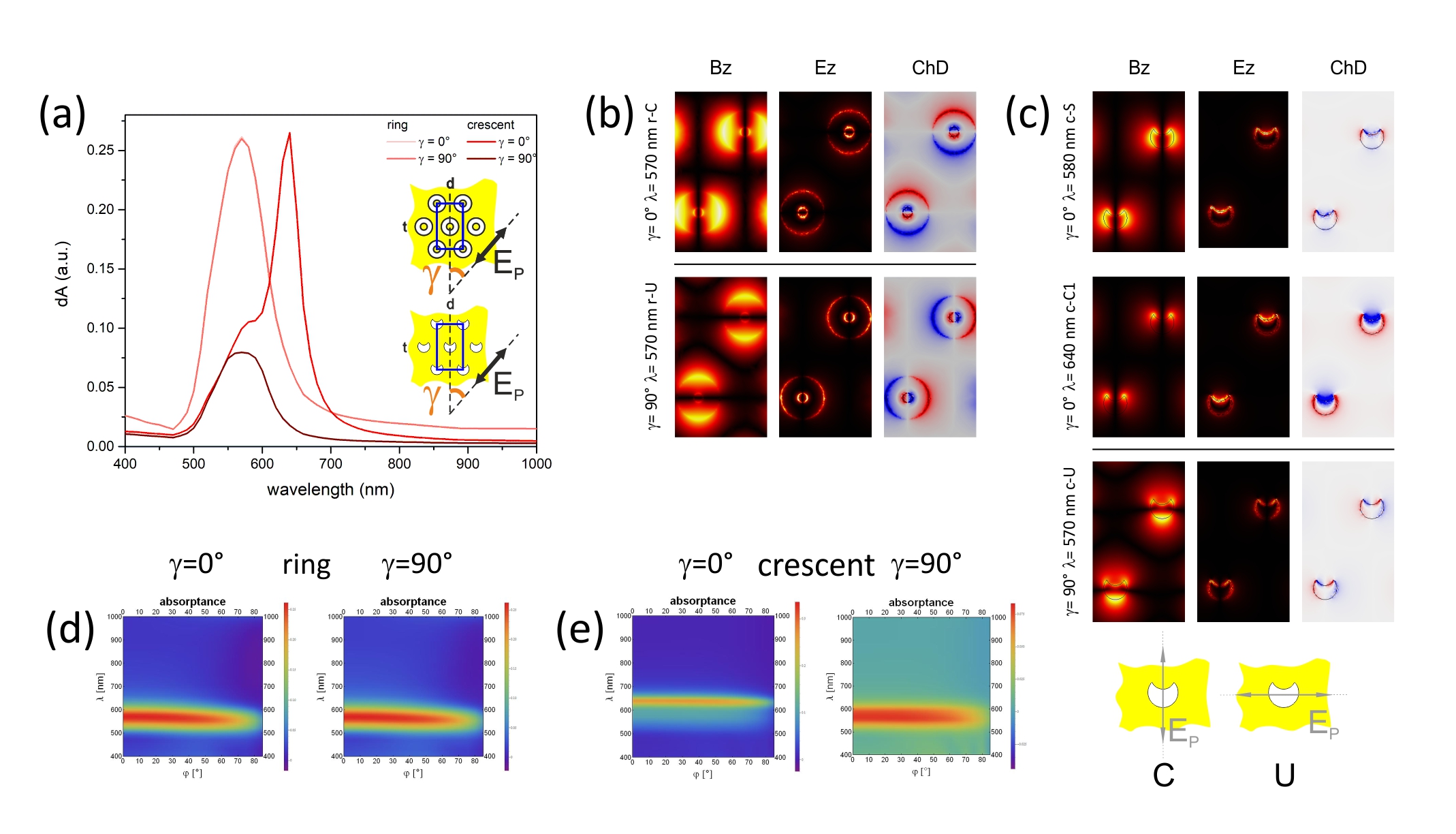}}
\caption{Hexagonal patterns composed of concave nanorings and horizontal nanocrescents: (a) absorptance spectra, (b, c) $B_z$ - $E_z$ field component and charge distribution in (top) $0^{\circ}$ and (bottom) $90^{\circ}$ azimuthal orientation of (b) nanorings and (c) nanocrescents, (d, e) dispersion characteristics taken in (left) $0^{\circ}$ and (right) $90^{\circ}$ azimuthal orientation of (d) nanorings and (e) nanocrescents. Insets: schematic drawing of the unit cells showing the C and U configurations.}
\label{IICSM_DG_cv_01}
\end{figure}

Due to the spherical symmetry of the composing nano-objects and to the symmetry properties of the hexagonal lattice, on the rectified absorptance of the hexagonal pattern composed of concave nanorings one single maximum appears at 570 nm azimuthal orientation independently (Fig. \ref{IICSM_DG_cv_01}a). Moreover, the global maxima completely overlap in $90^{\circ}$ and $0^{\circ}$ azimuthal orientations. The $B_z$ field component distribution indicates lobes both on the inner and outer rim of the concave nanorings perpendicularly to the $\bar{E}$-field oscillation direction, which corresponds with the $E_z$ field component distribution lobes on the complementary hexagonal pattern of convex nanorings \cite{ref_49,ref_50}. In contrast, the accompanying $E_z$ field component distribution indicates lobes parallel to the $\bar{E}$-field oscillation direction on the concave nanorings, which promotes to uncover the characteristic charge distribution consisting of reversal dipoles on the inner and outer rim of the nanorings at the indistinguishable r-C and r-U resonance (Fig. \ref{IICSM_DG_cv_01}b, top and bottom).

\subsubsection{Hexagonal pattern of horizontal concave nanocrescents}
\label{Hex_cre}

\paragraph{Hexagonal pattern of horizontal concave nanocrescents in C orientation}
\label{Hex_cre_C}
 \ \\ 
On the rectified absorptance of the hexagonal array composed of horizontal concave nanocrescents a shoulder (580 nm) is followed by a global maximum (640 nm) in $0^{\circ}$ azimuthal orientation, which is the C orientation (Fig. \ref{IICSM_DG_cv_01}a). On the $B_z$ field component distribution the neighbouring four c-C2 resonance related and two c-C1 resonance related lobes result in two well separated composite lobes at the shoulder. This is accompanied by three bright and three weak lobes on the $E_z$ field component distribution on the small and large arch of the nanocrescents, respectively. Accordingly, the characteristic charge distribution is hexapolar. At the global maximum the two tips are shiny on the $B_z$ field component distribution, that is accompanied by two lobes of the $E_z$ field component distribution on the archs of the nanocrescents, which is significantly stronger on the smaller arch. The characteristic charge distribution is dipolar along the $\bar{E}$-field oscillation direction at the c-C1 resonance (Fig. \ref{IICSM_DG_cv_01}c, top).

\paragraph{Hexagonal pattern of horizontal concave nanocrescents in U orientation}
\label{Hex_cre_U}
 \ \\ 
In comparison, on the rectified absorptance of the hexagonal pattern composed of horizontal concave nanocrescents only the global maximum (570 nm) appears in $90^{\circ}$ azimuthal orientation, which is the U orientation (Fig. \ref{IICSM_DG_cv_01}a). At the global maximum there are two lobes, one on the large arch and the other distributed on the tips on the $B_z$ field component distribution, which is accompanied by four lobes of the $E_z$ field component distribution on the nanocrescents. The characteristic charge distribution is quadrupolar governed by the $\bar{E}$-field oscillation direction at the c-U resonance (Fig. \ref{IICSM_DG_cv_01}c, bottom).

\subsubsection{Rectangular 300 nm periodic pattern of singlet concave nanorings}
\label{Rec_rin}

\begin{figure}[h]
\center
\switchocg{imgA02 imgB02}{
  \makebox[0pt][l]{
    \begin{ocg}{Image A02}{imgA02}{on}
      \includegraphics[scale=0.6]{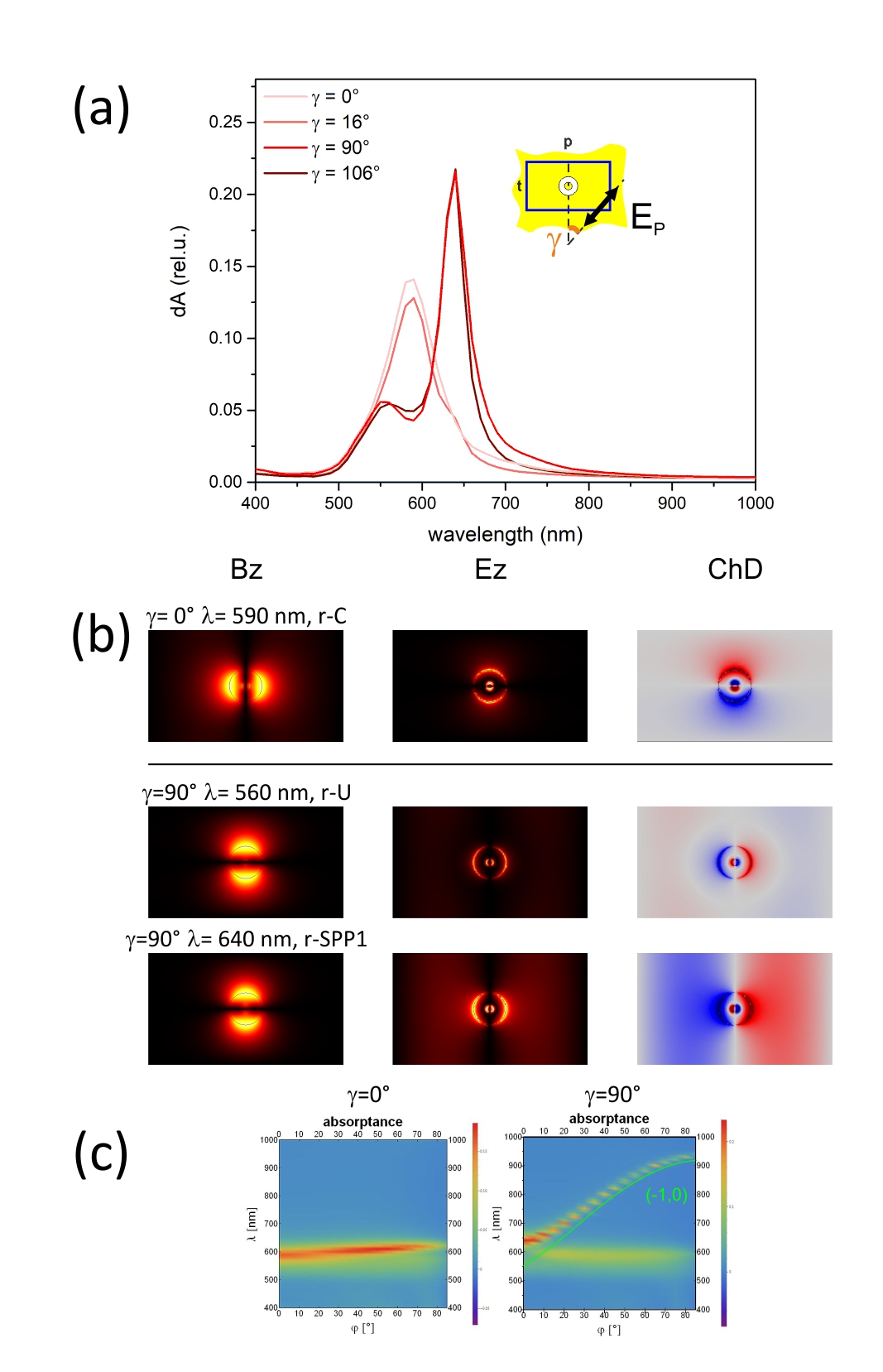}
    \end{ocg}
  }
  \begin{ocg}{Image B02}{imgB02}{off}
     \includegraphics[scale=0.6]{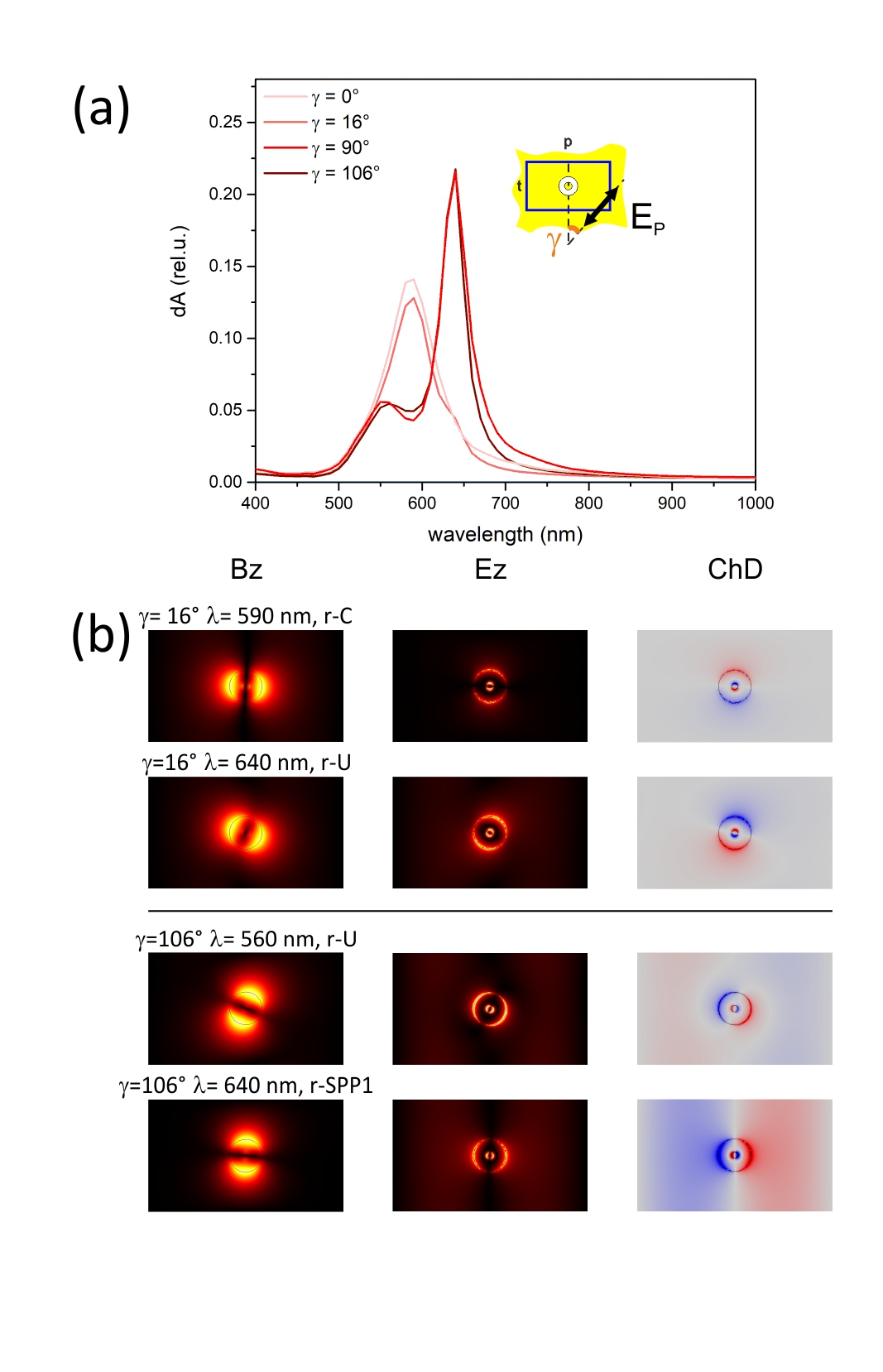}
  \end{ocg}
}
\caption{Rectangular p=300 nm periodic pattern composed of singlet concave nanorings: (a) absorptance spectra, (b) $B_z$ - $E_z$ field component and charge distribution in (top) $0^{\circ}$/$16^{\circ}$ and (bottom) $90^{\circ}$/$106^{\circ}$ azimuthal orientation, (c) dispersion characteristics taken in (left) $0^{\circ}$ and (right) $90^{\circ}$ azimuthal orientation. Inset: schematic drawing of the unit cell. (Please note that by clicking on the figure you can switch between $0^{\circ}$ and $16^{\circ}$ as well as $90^{\circ}$ and $106^{\circ}$.)}
\label{IICSM_DG_cv_02}
\end{figure}

\paragraph{Rectangular 300 nm periodic pattern of singlet concave nanorings in C orientation of nanocrescents}
\label{Rec_rin_C}
 \ \\ 
When singlet concave nanorings similar to those inspected in (i) hexagonal pattern are arranged into 300 nm periodic rectangular pattern, on the rectified absorptance a global maximum (590 nm) appears in $0^{\circ}$/$16^{\circ}$ azimuthal orientation, which is the C orientation of horizontal singlet / slightly rotated quadrumer concave nanocrescents, and is followed by a tiny shoulder (640 nm) in $16^{\circ}$ azimuthal orientation (Fig. \ref{IICSM_DG_cv_02}a). The $B_z$ ($E_z$) field component exhibits lobes perpendicularly (parallel) to the $\bar{E}$-field oscillation direction at the global maximum. Accordingly, the characteristic charge distribution at the r-C resonance consists of reversal dipoles on the inner and outer rim of the nanoring along the $\bar{E}$-field oscillation direction. This characteristic charge distribution is expected to be insensitive to the $\bar{E}$-field oscillation direction due to the spherical symmetry of the concave singlet nanorings. The tiny shoulder at $16^{\circ}$ azimuthal orientation indicates a cross-coupling effect due to existing $\bar{E}$-field component perpendicularly to the nanocrescent symmetry axis. However, instead of a horizontal cross-coupled r-U mode, the reversal dipoles rotate on the nanoring (Fig. \ref{IICSM_DG_cv_02}b, top).

\paragraph{Rectangular 300 nm periodic pattern of singlet concave nanorings in U orientation of nanocrescents}
\label{Rec_rin_U}
 \ \\ 
In comparison, on the rectified absorptance of 300 nm periodic rectangular array composed of singlet concave nanorings the local maximum appearing at a slightly smaller wavelength (560 nm / 560 nm) is followed by a global maximum (640 nm / 640 nm) in $90^{\circ}$/$106^{\circ}$ azimuthal orientation, which is the U orientation of horizontal singlet / slightly rotated quadrumer concave nanocrescents (Fig. \ref{IICSM_DG_cv_02}a). At the local maximum the $B_z$ ($E_z$) field component lobes are perpendicular (parallel) to the $\bar{E}$-field oscillation direction. Accordingly, the characteristic charge distribution at the r-U resonance consists of reversal dipoles on the inner and outer rim on the concave singlet nanorings along the $\bar{E}$-field oscillation direction. In contrast, at the global maximum in $90^{\circ}$ / $106^{\circ}$ azimuthal orientation the $B_z$ ($E_z$) field lobes are completely perpendicular (parallel) to the $\bar{k}_p$ lattice vectors / are rotated clockwise with a smaller extent compared to the local maximum, which indicates a dominance of a grating-coupling effect with a strength correlating with the $\bar{E}$-field component along the $\bar{k}_p$ direction. Accordingly, at the global maximum (also in case of $106^{\circ}$ azimuthal orientation) the horizontal reversal dipoles and the strong periodic charge modulation along the $\bar{k}_p$ lattice vector prove the grating-coupling of SPP1 modes in (-1, 0) order, which is more / less pronounced in $90^{\circ}$/$106^{\circ}$ azimuthal orientation (Fig. \ref{IICSM_DG_cv_02}b, bottom and 4c right).

\subsubsection{Rectangular 300 nm periodic pattern of singlet horizontal concave nanocrescents}
\label{Rec_cre}

\begin{figure}[h]
\center
\switchocg{imgA03 imgB03}{
  \makebox[0pt][l]{
    \begin{ocg}{Image A03}{imgA03}{on}
      \includegraphics[scale=0.6]{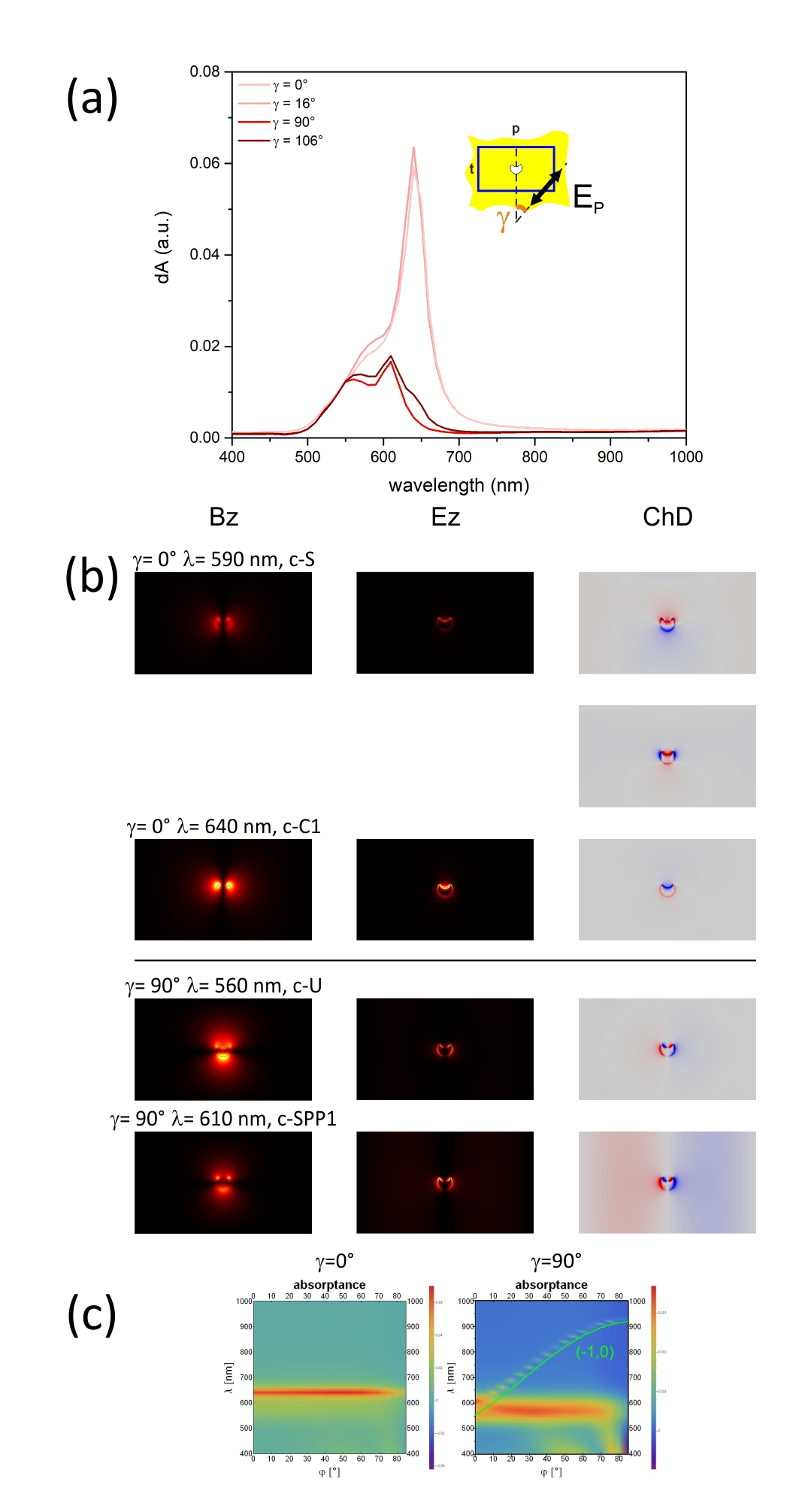}
    \end{ocg}
  }
  \begin{ocg}{Image B03}{imgB03}{off}
     \includegraphics[scale=0.6]{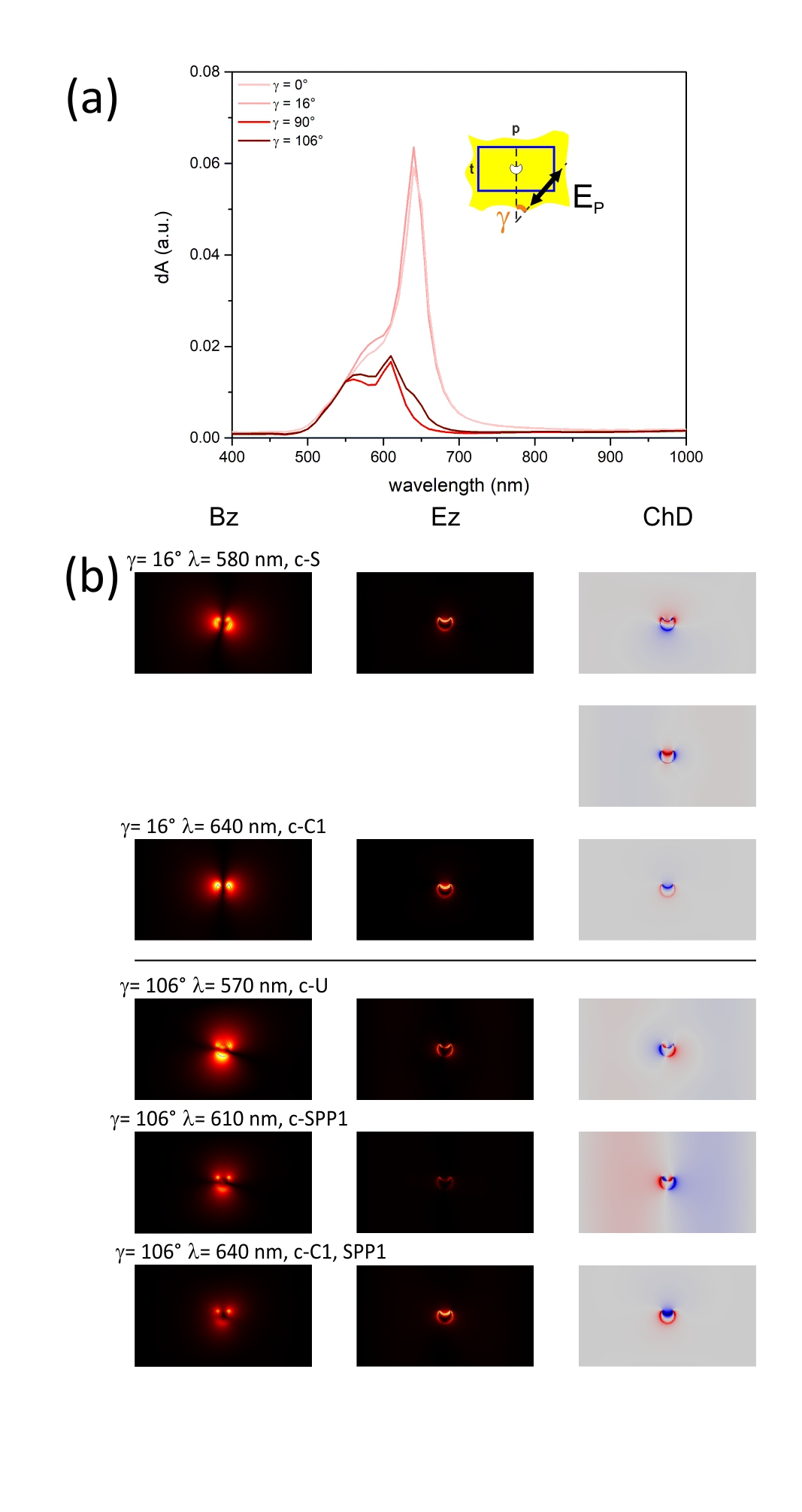}
  \end{ocg}
}
\caption{Rectangular p=300 nm periodic pattern composed of horizontal singlet concave nanocrescents: (a) absorptance spectra, (b) $B_z$- $E_z$ field component and charge distribution in (top) $0^{\circ}$/$16^{\circ}$ and (bottom) $90^{\circ}$/$106^{\circ}$ azimuthal orientation, (c) dispersion characteristics taken in (left) $0^{\circ}$ and (right) $90^{\circ}$ azimuthal orientation. Inset: schematic drawing of the unit cell. (Please note that by clicking on the figure you can switch between $0^{\circ}$ and $16^{\circ}$ as well as $90^{\circ}$ and $106^{\circ}$.)}
\label{IICSM_DG_cv_03}
\end{figure}

\paragraph{Rectangular 300 nm periodic pattern of singlet horizontal concave nanocrescent in C orientation}
\label{Rec_cre_C}
 \ \\ 
When singlet horizontal concave nanocrescents are arranged into 300 nm periodic rectangular array, on the rectified absorptance only a shoulder appears (590 nm / 580 nm) before the global maximum (640 nm / 640 nm) in C orientation ($0^{\circ}$) / close to it ($16^{\circ}$) (Fig. \ref{IICSM_DG_cv_03}a). On the $B_z$ field component the closely neighbouring four c-C2 resonance related and two c-C1 resonance related lobes results in four lobes at the shoulder, with intensity maxima on the tips. This is accompanied by two $E_z$ field component lobes, the stronger one appears on the smaller arch of the nanocrescents. A quadrupolar and hexapolar charge distribution is also observable in a noticeable fraction within one-cycle of the time-dependent charge distribution. In contrast, at the global maximum the two tips are shiny on the $B_z$ field component distribution, whereas the accompanying $E_z$ field component indicates two lobes on the two archs of the nanocrescents, the stronger lobe appears on the smaller arch both in $0^{\circ}$ and $16^{\circ}$ azimuthal orientations. The characteristic charge distribution is dipolar along the $\bar{E}$-field oscillation direction at the c-C1 resonance on the singlet concave nanocrescent in $0^{\circ}$ azimuthal orientation (Fig. \ref{IICSM_DG_cv_03}b, top).

\paragraph{Rectangular 300 nm periodic pattern of singlet horizontal concave nanocrescents in U orientation}
\label{Rec_cre_U}
 \ \\ 
In comparison, on the rectified absorptance of the 300 nm periodic rectangular array composed of concave nanocrescents a local maximum (560 nm / 570 nm) appears before the global maximum (610 nm / 610 nm) in U orientation ($90^{\circ}$) / close to it ($106^{\circ}$). In addition to this there is a shoulder at 640 nm in $106^{\circ}$ azimuthal orientation (Fig. \ref{IICSM_DG_cv_03}a). The $B_z$ field component exhibits two lobes on the archs of the nanocrescents at the local maximum, the stronger lobe appears on the larger arch. This is accompanied by $E_z$ field component exhibiting four lobes on the nanocrescent. Accordingly, a quadrupolar charge distribution is characteristic at the c-U resonance on the horizontal singlet concave nanocrescent (Fig. \ref{IICSM_DG_cv_03}b, bottom). \\
In contrast, at the global maximum from the two lobes the one distributed on the tips becomes commensurately shiny with that on the larger arch (the bridge between the tips disappears) on the $B_z$ field component distribution, whereas the accompanying $E_z$ field component indicates still four lobes, but two become more shiny on the smaller arch close to the tips of the nanocrescent. The characteristic charge distribution is still quadrupolar. In addition to this stronger/weaker periodic charge modulation appears as well, which originates from grating-coupling of SPP1 modes in (-1, 0) order. This reveals that SPPs are more / less efficiently coupled due to the $\bar{E}$-field oscillation direction along / almost parallel to $\bar{k}_p$ in $90^{\circ}$/ $106^{\circ}$ azimuthal orientation (Fig. \ref{IICSM_DG_cv_03}b, bottom and 5c right). \\
At the shoulder appearing exclusively in $106^{\circ}$ azimuthal orientation on the $B_z$ field component two lobes are observable, the one distributed on the tips is asymmetric. This is accompanied by two lobes on the $E_z$ field component distribution on the nanocrescent archs, with larger intensity on the small arch. The dominant charge distribution indicates the dipolar c-C1 resonance, which is cross-coupled due to the $\bar{E}$-field component along the symmetry axis of the nanocrescent (Fig. \ref{IICSM_DG_cv_03}b, bottom). \\
The spectrum in $106^{\circ}$ azimuthal orientation is unique, since the c-U resonance (570 nm), is followed by a peak caused by SPP1 grating-coupling (610 nm), than a shoulder (640 nm) originating from cross-coupled c-C1 appears.

\subsubsection{Rectangular 300 nm periodic pattern of quadrumer concave nanocrescents}
\label{Rec_qua}

\begin{figure}[h]
\center
\switchocg{imgA04 imgB04}{
  \makebox[0pt][l]{
    \begin{ocg}{Image A04}{imgA04}{on}
      \includegraphics[scale=0.6]{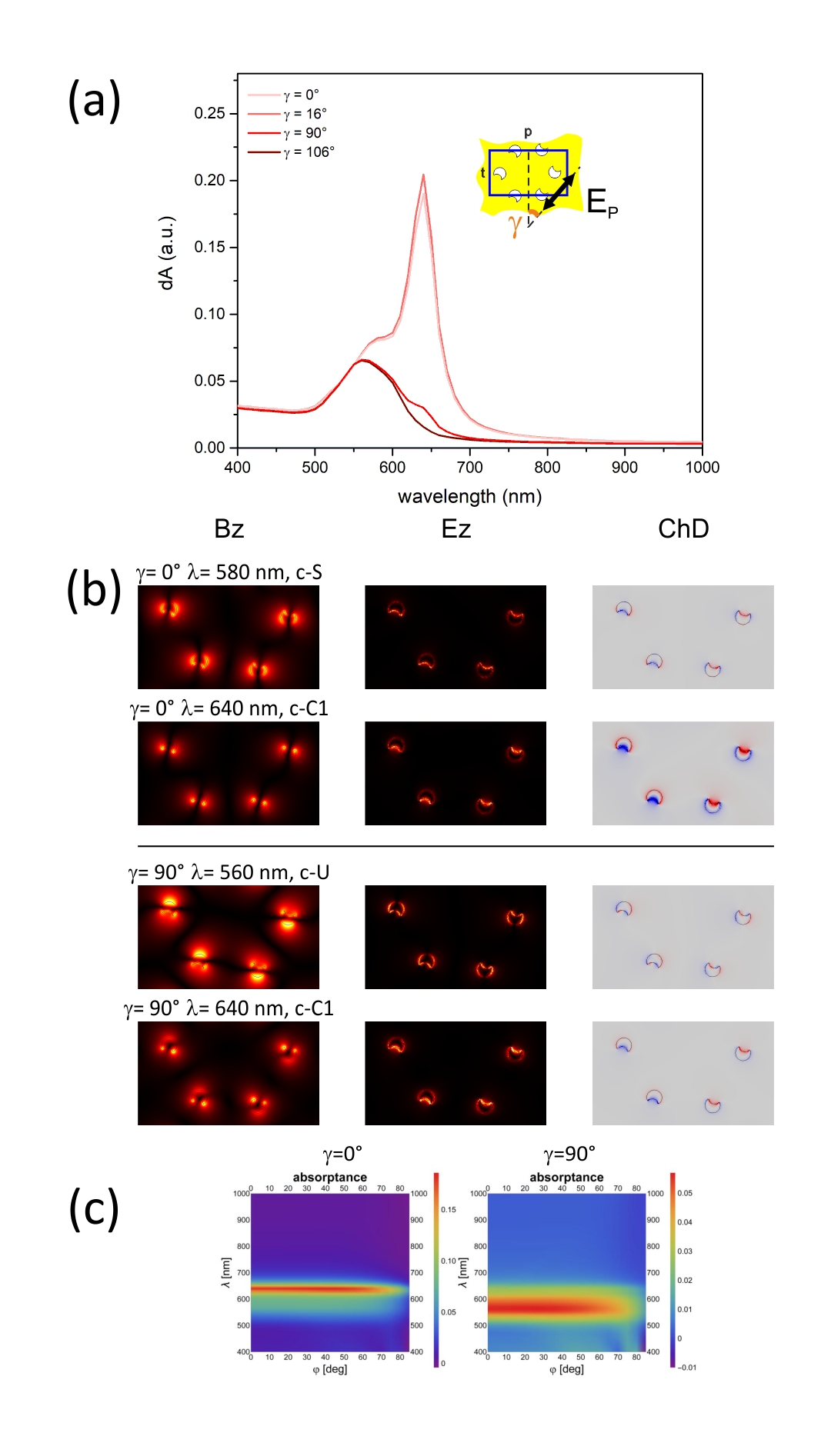}
    \end{ocg}
  }
  \begin{ocg}{Image B04}{imgB04}{off}
     \includegraphics[scale=0.6]{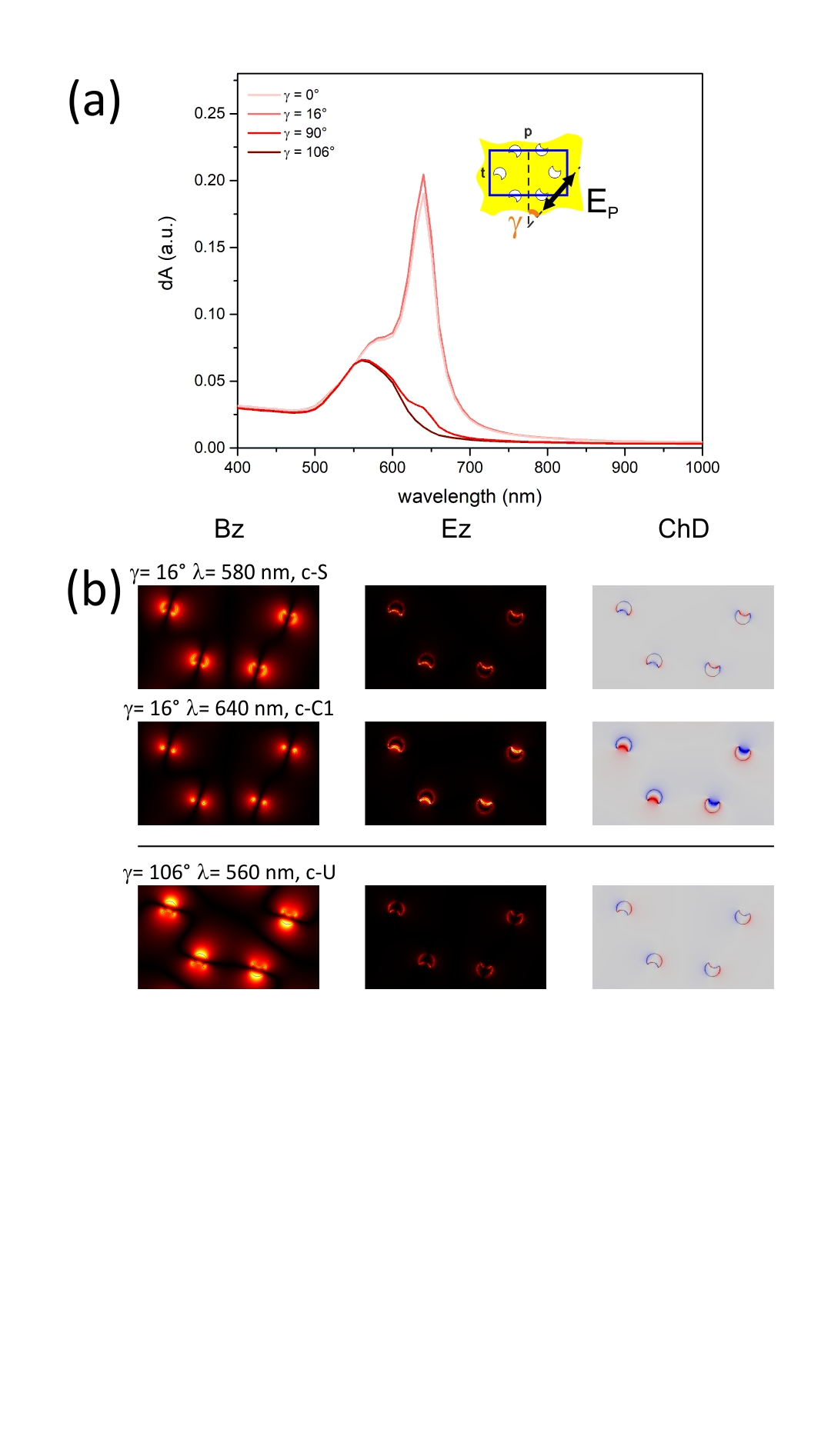}
  \end{ocg}
}
\caption{Rectangular p=300 nm periodic pattern composed of quadrumer concave nanocrescents: (a) absorptance spectra, (b) $B_z$-$E_z$ field component and charge distribution in (top) $16^{\circ}$/$0^{\circ}$ and (bottom) $106^{\circ}$/$90^{\circ}$ azimuthal orientation, (c) dispersion characteristics taken in (left) $0^{\circ}$ and (right) $90^{\circ}$ azimuthal orientation. Insets: schematic drawing of the unit cell. (Please note that by clicking on the figure you can switch between $0^{\circ}$ and $16^{\circ}$ as well as $90^{\circ}$ and $106^{\circ}$.)}
\label{IICSM_DG_cv_04}
\end{figure}

\paragraph{Rectangular 300 nm periodic pattern of quadrumer concave nanocrescents in C orientation}
\label{Rec_qua_C}
 \ \\ 
When 300 nm rectangular pattern is composed of four slightly rotated concave nanocrescents on the rectified absorptance a shoulder appears (580 nm / 580 nm) before the global maximum (640 nm /640 nm) close to ($0^{\circ}$) / in C orientation ($16^{\circ}$) of the quadrumers (Fig. \ref{IICSM_DG_cv_04}a). In contrast to the convex quadrumer counterpart, in case of concave patterns there is no difference between the extrema observable in presence of either one or four nanocrescents \cite{ref_49,ref_50}. The field distributions are analogous with those observable on singlet nanocrescents. Namely, caused by the coalescence of four c-C2 resonance related and two c-C1 resonance related lobes the $B_z$ field component exhibits four lobes at the shoulder, as a result intensity maxima appear on the tips. This is accompanied by two $E_z$ field component lobes on the nanocrescent archs, the stronger one appears on the smaller arch. Mainly quadrupolar charge distribution is observable, i.e. there is no hexapolar modulation in contrast to the horizontal singlet nanocrescent case. At the global maximum two lobes located on the nanocrescent tips are shiny on the $B_z$ field component distribution, whereas the accompanying $E_z$ field component distribution indicates two lobes on the archs of the nanocrescents, the lobe on the smaller arch is significantly stronger. The characteristic charge distribution is dipolar along the $\bar{E}$-field oscillation direction at the c-C1 resonance on the quadrumer of four nanocrescents in $16^{\circ}$ azimuthal orientation (Fig. \ref{IICSM_DG_cv_04}b, top).  

\paragraph{Rectangular 300 nm periodic pattern of quadrumer concave nanocrescents in U orientation}
\label{Rec_qua_U}
 \ \\ 
In comparison, on the rectified absorptance of the 300 nm periodic rectangular array composed of concave nanocrescents the global maximum (560 nm) is followed by a shoulder (640 nm) in $90^{\circ}$ azimuthal orientation, whereas in $106^{\circ}$ azimuthal orientation, which is the U orientation, only a single maximum (560 nm) appears (Fig. \ref{IICSM_DG_cv_04}a). The most important difference with respect to singlet horizontal nanocrescent containing unit cell is the partial / complete disappearance of the second maximum in $90^{\circ}$ / $106^{\circ}$ orientation. The $B_z$ field component exhibits two lobes at the global maximum, the stronger lobes appear on the larger arch of the nanocrescents, whereas the weaker lobes are asymmetrically / symmetrically aligned on the tips in $90^{\circ}$ / $106^{\circ}$ orientation. The accompanying $E_z$ field component distribution indicates four lobes on the nanocrescents, which are asymmetrically/ symmetrically aligned with respect to their axis. Accordingly, a quadrupolar charge distribution is observable at the c-U resonance on the quadrumer of four nanocrescents in $106^{\circ}$ azimuthal orientation (Fig. \ref{IICSM_DG_cv_04}b, bottom). \\
In contrast, at the shoulder appearing exclusively in $90^{\circ}$ azimuthal orientation on the $B_z$ field component distribution from the co-existent two lobes the one on the tips / larger arch become more / less shiny, whereas the $E_z$ field distribution indicates only two lobes, the one on the small arch of the nanocrescents becomes more shiny. The characteristic charge distribution is dipolar, as a result the quadrumer of nanocrescents has a net dipole moment. Comparison with the charge distribution observable in C orientation at the same spectral position reveals that a dipolar charge distribution analogue with that of c-C1 appears. This is due to the $\bar{E}$-field component along the symmetry axes of the concave nanocrescents that enables cross-coupling of c-C1 resonance in $90^{\circ}$ azimuthal orientation (Fig. \ref{IICSM_DG_cv_04}b, bottom). Surprisingly there is no signature of SPP1 coupling caused by the group symmetry of the quadrumer (Fig. \ref{IICSM_DG_cv_04}c, right).

\subsubsection{Rectangular 300 nm periodic pattern of complex concave miniarray}
\label{Min_arr_300}

\begin{figure}[h]
\center
\switchocg{imgA05 imgB05}{
  \makebox[0pt][l]{
    \begin{ocg}{Image A05}{imgA05}{on}
      \includegraphics[scale=0.6]{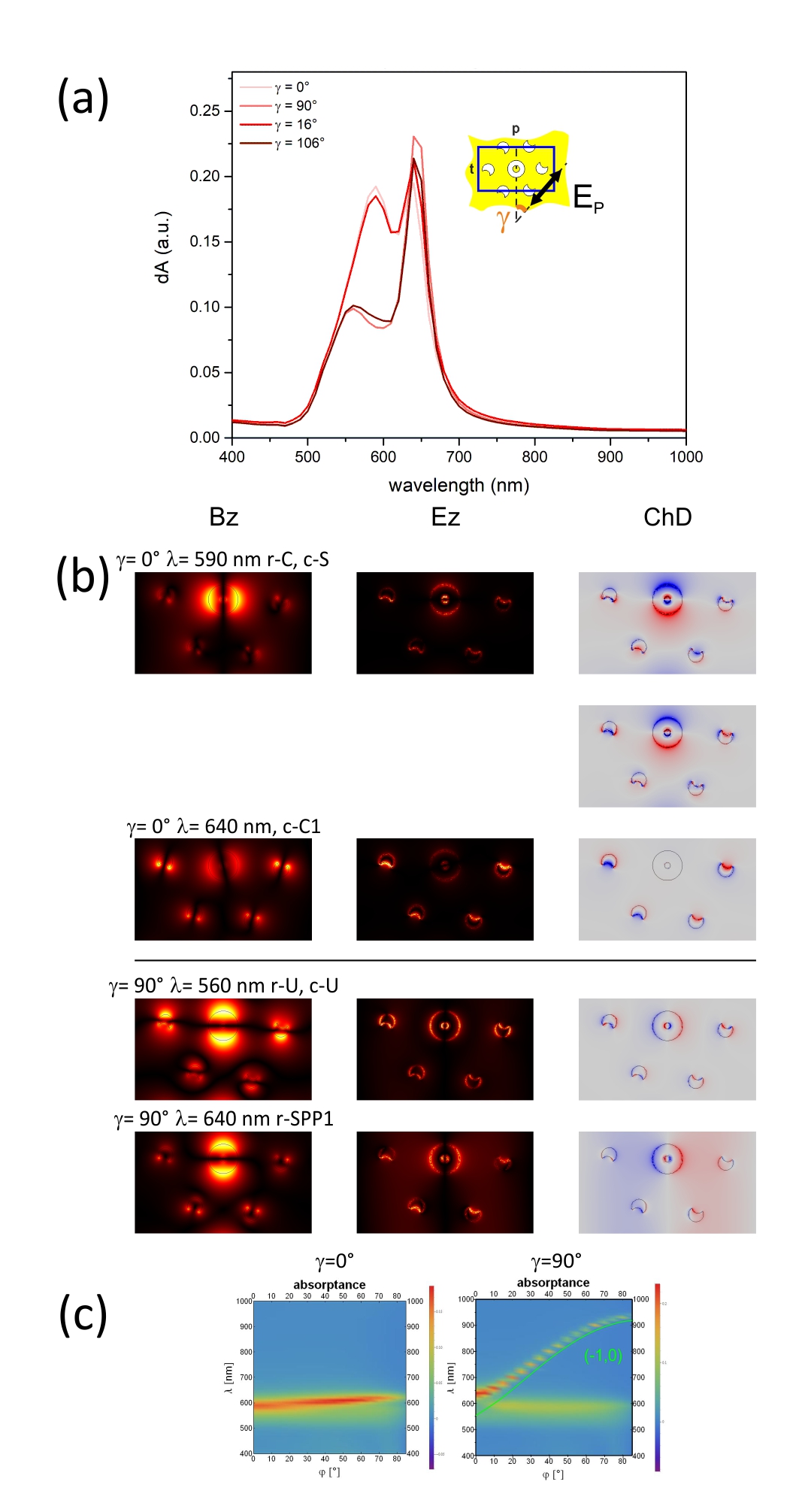}
    \end{ocg}
  }
  \begin{ocg}{Image B05}{imgB05}{off}
     \includegraphics[scale=0.6]{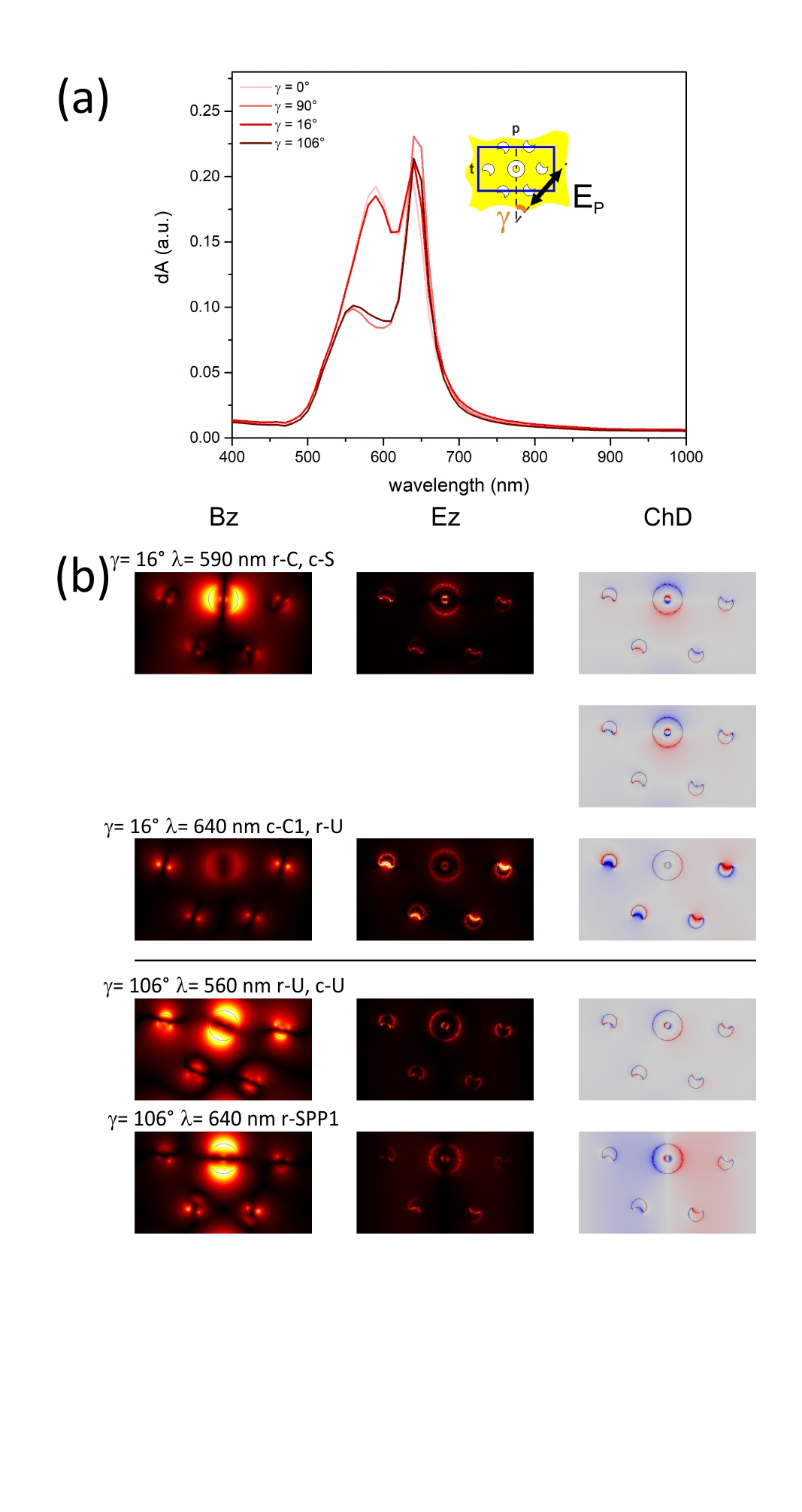}
  \end{ocg}
}
\caption{Rectangular p=300 nm periodic pattern composed of complex concave miniarray: (a) absorptance spectra, (b) $B_z$-$E_z$ field component and charge distribution in (top) $16^{\circ}$ / $0^{\circ}$ and (bottom) $106^{\circ}$ / $90^{\circ}$ azimuthal orientation, (c) dispersion characteristics taken in (b) $0^{\circ}$ and (c) $90^{\circ}$ azimuthal orientation. Inset: schematic drawing of the unit cell. (Please note that by clicking on the figure you can switch between $0^{\circ}$ and $16^{\circ}$ as well as $90^{\circ}$ and $106^{\circ}$.)}
\label{IICSM_DG_cv_05}
\end{figure}

\paragraph{Rectangular 300 nm periodic pattern of complex concave miniarray in C orientation}
\label{Min_arr_300_C}
 \ \\ 
When 300 nm periodic rectangular pattern is composed of complex concave miniarray consisting of both the central nanoring and quadrumer of nanocrescents, on their absorptance a local maximum appears (590 nm / 590 nm) before the global maximum (640 nm / 640 nm) close to ($0^{\circ}$) / in C orientation ($16^{\circ}$) (Fig. \ref{IICSM_DG_cv_05}a). There is only a slight difference between the local and global maximum of the complex miniarray and (global maximum) shoulder and (-/shoulder) global maximum observable in case of (singlet nanoring) quadrumer of four nanocrescents. The $B_z$ field component exhibits intense lobes on the outer rim of the nanoring at the local maximum perpendicularly to the $\bar{E}$-field oscillation direction. Significantly weaker four lobes appear on the nanocrescents caused by the coalescence of the four and two lobes originating from c-C2 and c-C1 resonance, as a result asymmetrical intensity maxima appear on the tips of the nanocrescents. This is accompanied by $E_z$ field component distribution, which indicates stronger lobes on the inner rim of the nanoring along the $\bar{E}$-field oscillation direction, and two lobes on the archs of the nanocrescents, the one on the smaller arch is stronger. Strong reversal dipoles arise on the inner and outer rim of the nanoring along the $\bar{E}$-field oscillation direction, which are enhanced especially on the inner rim. Not only quadrupolar charge distribution is observable, a hexapolar charge modulation also develops similarly to the singlet nanocrescent but in contrast to the quadrumer nanocrescent case (Fig. \ref{IICSM_DG_cv_05}b, top). \\
In contrast, at the global maximum the nanoring exhibits significantly weaker lobes, which appear exclusively on the outer rim, whereas the two lobes localized onto the nanocrescent tips are asymmetrically/symmetrically shiny on the $B_z$ field component distribution. Accordingly, the accompanying $E_z$ field component indicates weak lobes of less well-defined directivity on the nanoring, whereas significantly stronger two lobes appear on the archs of the nanocrescents, especially on their smaller arch. Dipolar charge distribution is develops along the $\bar{E}$-field oscillation direction on the nanocrescents, that exhibit pure c-C1 resonance on the quadrumer of them in $16^{\circ}$ azimuthal orientation, whereas weak reversal dipoles are rotating on the nanoring instead of an expectable horizontal cross-coupled r-U mode.

The local maximum on the miniarray originates from the r-C resonance of the nanoring, which overlaps with the interacting c-C2 and c-C1 modes on the nanocrescents in the quadrumer, the latter resulted in a shoulder on their spectrum as well. The global maximum originates mainly from the c-C1 mode on the nanocrescents quadrumer. The cross-coupled r-U resonance on the nanoring spectrally overlaps with the c-C1 resonance on the nanocrescents at the global maximum. The intermittent perpendicularity of the weak rotating nanoring dipole and strong nanocrescent dipoles results in a Fano interference between two LSPRs (Fig. \ref{IICSM_DG_cv_05}b, top).

\paragraph{Rectangular 300 nm periodic pattern of complex concave miniarray in U orientation}
\label{Min_arr_300_U}
 \ \\ 
In comparison, on the rectified absorptance of the 300 nm periodic rectangular array composed of complex concave miniarray the local maximum (560 nm / 560 nm) is followed by a narrower global maximum (640 nm / 640 nm) close to ($90^{\circ}$) / in U orientation ($106^{\circ}$) (Fig. \ref{IICSM_DG_cv_05}a). The most important difference with respect to the nanocrescent quadrumer containing unit cell is the recovery of the second maximum, which has been observed already in case of the singlet nanoring and singlet nanocrescent containing unit cells in $90^{\circ}$ azimuthal orientation. The $B_z$ field component exhibits strong lobes on the outer rim of the nanoring perpendicularly to the $\bar{E}$-field oscillation direction, and two lobes on the nanocrescents at the local maximum, the stronger lobe appears on the larger arch of the nanocrescents, whereas the weaker is asymmetrically / symmetrically distributed on the tips in $90^{\circ}$ / $106^{\circ}$ azimuthal orientation. This is accompanied by $E_z$ field component distribution indicating lobes along the $\bar{E}$-field oscillation direction on the inner and outer rim of the nanoring, and four lobes, which are asymmetrically/symmetrically aligned on the nanocrescents. Accordingly, reversal dipoles arise on the inner and outer rim of the nanoring along the $\bar{E}$-field oscillation direction at the r-U resonance, and quadrupolar charge distribution is observable on the nanocrescents at the c-U resonance (Fig. \ref{IICSM_DG_cv_05}b, bottom). In contrast, at the global maximum the two lobes on the $B_z$ field component distribution on the outer rim of the nanoring are aligned perpendicularly to the $\bar{k}_p$ vector both in $90^{\circ}$ and $106^{\circ}$ azimuthal orientation, whereas the asymmetrically distributed lobes on the tips become more shiny than the lobe on the larger arch of the nanocrescents. However, the contribution of nanocrescents in quadrumer is not significant at this extremum, as it is shown in Fig. \ref{IICSM_DG_cv_05}a and b, bottom. The accompanying $E_z$ field component distribution indicates two lobes on the nanoring almost along the $\bar{k}_p$ vector, and two asymmetrical lobes that are more shiny on the small arch of the nanocrescents. The characteristic charge distribution includes reversal dipolar distribution on the inner and outer rim on the nanoring along / slightly rotated with respect to the $\bar{k}_p$ vector, and a weak quadrupolar distribution on the nanocrescents. The extended periodic charge modulation corresponds to SPP1 grating-coupled in (-1,0) order ($k_p$ $k_{SPP}$). The miniarray local maximum originates from the sum of the coincident r-U mode of the nanoring and c-U modes on the quadrumer of nanocrescents, whereas the grating-coupled SPP1 is responsible for the narrow global maximum (Fig. \ref{IICSM_DG_cv_05}b, bottom and \ref{IICSM_DG_cv_05}c, right).

\subsubsection{Rectangular 600 nm periodic pattern of complex concave miniarray}
\label{Min_arr_600}

\begin{figure}[h]
\center
\switchocg{imgA06 imgB06}{
  \makebox[0pt][l]{
    \begin{ocg}{Image A06}{imgA06}{on}
      \includegraphics[scale=0.6]{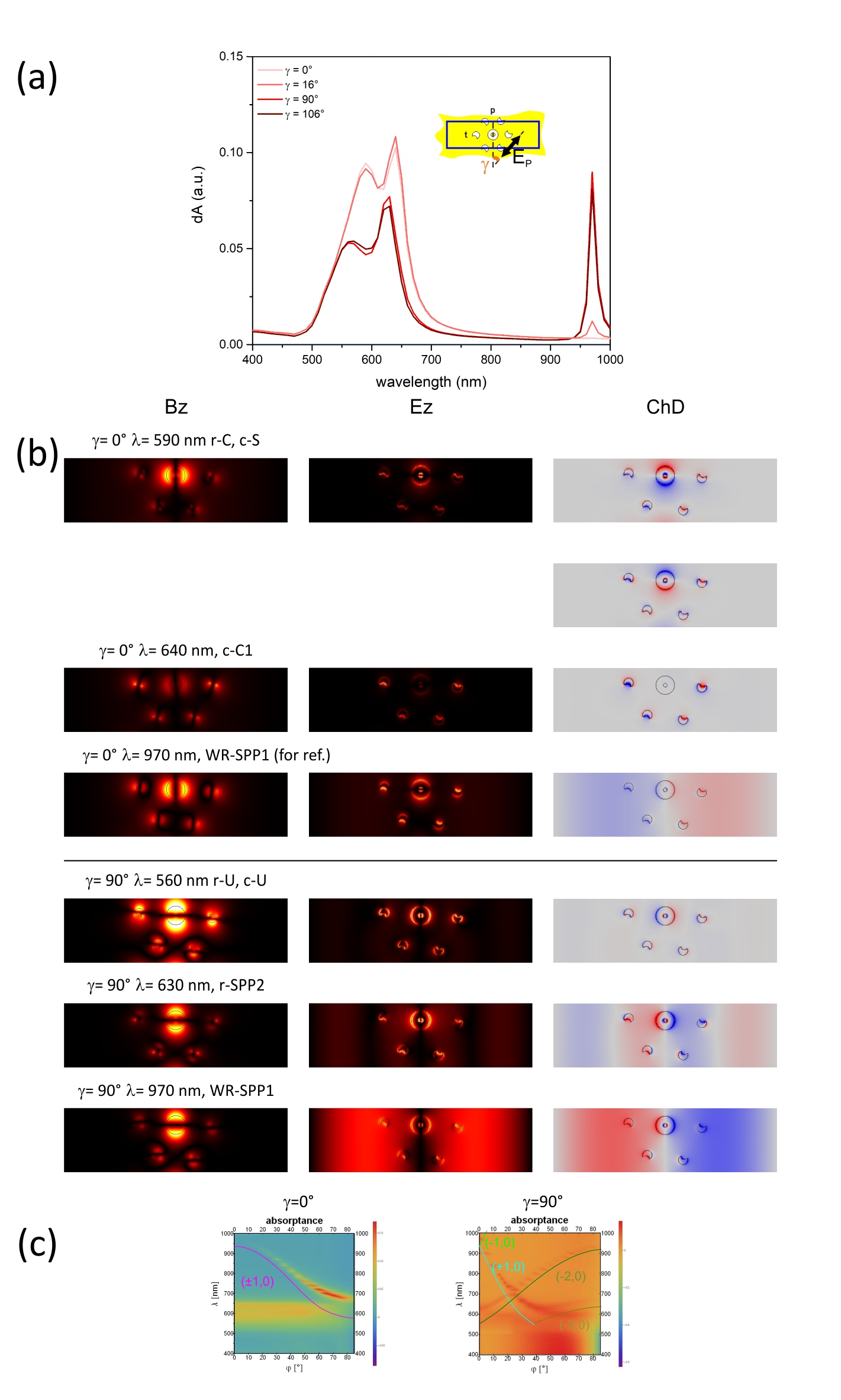}
    \end{ocg}
  }
  \begin{ocg}{Image B06}{imgB06}{off}
     \includegraphics[scale=0.6]{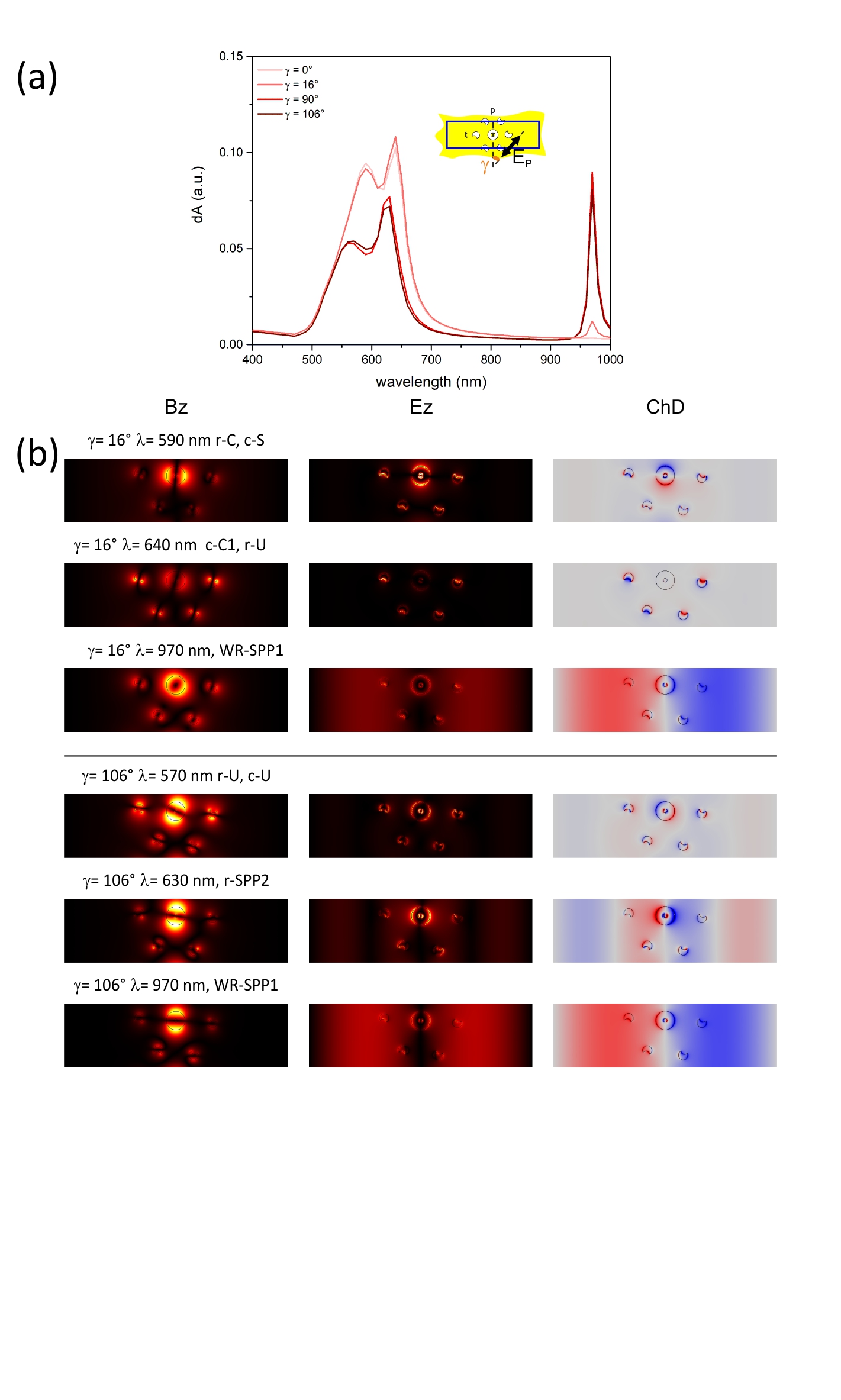}
  \end{ocg}
}
\caption{Rectangular p=600 nm periodic pattern composed of complex concave miniarray: (a) absorptance spectra, (b) $B_z$-$E_z$ field component and charge distribution in (top) $16^{\circ}$ / $0^{\circ}$ and (bottom) $106^{\circ}$ / $90^{\circ}$ azimuthal orientation, (c) dispersion characteristics taken in (b) $0^{\circ}$ and (c) $90^{\circ}$ azimuthal orientation. Inset: schematic drawing of the unit cell. (Please note that by clicking on the figure you can switch between $0^{\circ}$ and $16^{\circ}$ as well as $90^{\circ}$ and $106^{\circ}$.)}
\label{IICSM_DG_cv_06}
\end{figure}

\paragraph{Rectangular 600 nm periodic pattern of complex concave miniarray in C orientation}
\label{Min_arr_600_C}
 \ \\ 
When two-times larger 600 nm periodic rectangular pattern is composed of the same complex concave miniarray consisting of the central nanoring and quadrumer of nanocrescents, on their absorptance a local maximum appears (590 nm / 590 nm) before the global maximum (640 nm / 640 nm), moreover no / a significant modulation appears at larger wavelength (- / 970 nm) close to ($0^{\circ}$) / in C orientation ($16^{\circ}$) (Fig. \ref{IICSM_DG_cv_06}a). Similarly to the 300 nm periodic rectangular pattern there is no (only a slight 10 nm) difference between the local maximum of the complex miniarray and the global maximum (shoulder) observable in case of a singlet nanoring (quadrumer of nanocrescents). The $B_z$ field component exhibits lobes on the outer rim of the nanoring at the local maximum perpendicularly to the $\bar{E}$-field oscillation direction. Significantly weaker four $B_z$ lobes appear on the nanocrescents caused by the coalescence of the four and two lobes originating from c-C2 and c-C1 resonance, with significantly/considerably asymmetrical distribution, which is noticeable mainly on the tips on the nanocrescents. This is accompanied by $E_z$ field component distribution, which indicates stronger lobes on the inner rim of the nanoring along the $\bar{E}$-field oscillation direction, and two lobes on the nanocrescents, the stronger lobe appears on the smaller arch of them. Strong reversal dipoles arise on the inner and outer rim of the nanoring along the $\bar{E}$-field oscillation direction at the r-C resonance. Mainly quadrupolar charge distribution is observable, but there is a hexapolar modulation on the components of the quadrumer as well, similarly to the charge distribution at the shoulder on singlet nanocrescents.

Similarly to the 300 nm periodic rectangular pattern there is no difference between the global maximum of the complex miniarray and the (- / shoulder) global maximum observable in case of a (singlet nanoring) quadrumer of nanocrescents. Compared to the local maximum, at the global maximum the nanoring exhibits significantly weaker lobes on the outer rim perpendicularly to the $\bar{E}$-field oscillation direction, whereas the two tips are asymmetrically / symmetrically shiny on the $B_z$ field component distribution. The accompanying $E_z$ field component distribution indicates two lobes, the stronger one appears on the smaller arch of the nanocrescents. The charge accumulation is almost negligible on the nanoring, whereas it is dipolar along the $\bar{E}$-field oscillation direction on the nanocrescents at the c-C1 resonance on the quadrumer in $16^{\circ}$ azimuthal orientation of the miniarray (Fig. \ref{IICSM_DG_cv_06}b, top). The miniarray response originates again from the sum of the r-C mode of the nanoring overlapping with interacting c-C2 and c-C1 modes at the local maximum, and from the c-C1 modes on the nanocrescents in quadrumer at the global maximum. In addition to this the r-U resonance on the nanoring cross-coupled in $16^{\circ}$ azimuthal orientation spectrally overlaps with the c-C1 resonance on the nanocrescents at the global maximum, similarly to the 300 nm periodic pattern of the miniarray.\\
The local maximum appearing at larger wavelength in $16^{\circ}$ azimuthal orientation is exceptional for the larger periodic rectangular pattern. Here the $B_z$($E_z$) field component aligned perpendicularly (parallel) to the $\bar{E}$-field oscillation direction is more/less well defined in $0^{\circ}$/$16^{\circ}$, in addition to this there is a periodic lateral modulation on $E_z$ field component in $16^{\circ}$, which is significantly weaker at the same wavelength in $0^{\circ}$ azimuthal orientation, i.e. the latter is shown as a reference. The charge distribution consists of reversal dipoles on the nanoring along the $\bar{k}_p$ vector, and dipoles on the nanocrescents parallel to that on the outer rim of the nanoring. A lateral charge modulation with a period commensurate with the grating-coupled SPP1 wavelength appears as well. The preference to $16^{\circ}$ azimuthal orientation indicates that existence of an $\bar{E}$-field component along the $\bar{k}_p$ vector is required, and proves that a propagating plasmonic rather than a scattered photonic mode is at play. Indeed, the wavelength of SPP1 that can be coupled in ($\pm$1,0) order at this wavelength is closer to 600 nm period. At this Wood-Rayleigh anomaly the $E_z$ lobes are less perfectly oriented along the $\bar{E}$-field oscillation direction on the nanoholes, since the charge separation orientation is governed by the SPP1 having a wave vector commensurate with the $\bar{k}_p$ vector (Fig. \ref{IICSM_DG_cv_06}b, top and \ref{IICSM_DG_cv_06}c, left).

\paragraph{Rectangular 600 nm periodic pattern of complex concave miniarray in U orientation}
\label{Min_arr_600_U}
 \ \\ 
In comparison, on the rectified absorptance of the 600 nm periodic rectangular array composed of complex concave miniarray the local maximum (560 nm / 570 nm) is followed by a large global maximum (630 nm / 630 nm), and a huge modulation appears at larger wavelength (970 nm / 970 nm) close to ($90^{\circ}$) / in U orientation ($106^{\circ}$) (Fig. \ref{IICSM_DG_cv_06}a). The recovery of the second maximum that has been observed in case of the singlet nanoring and singlet nanocrescent containing unit cells is observable similarly to the 300 nm periodic pattern. At the local maximum the $B_z$ field component exhibits strong lobes on the outer rim of the nanoring perpendicularly to the $\bar{E}$-field oscillation direction, and two lobes on the nanocrescents, the stronger lobe appears on the larger arch of the nanocrescents, whereas the weaker is distributed asymmetrically / symmetrically on the tips in $90^{\circ}$/$106^{\circ}$ orientation. The accompanying $E_z$ field component distribution indicates lobes along the $\bar{E}$-field oscillation direction on the inner and outer rim of the nanoring, and four lobes, which are asymmetrically / symmetrically aligned on the nanocrescents. Accordingly, reversal dipoles develop on the inner and outer rim of the nanoring along the $\bar{E}$-field oscillation direction, and a quadrupolar charge distribution is observable on the nanocrescents, which corresponds to the r-U and c-U resonance on the complex concave miniarray. At the global maximum the lobes on the $B_z$ field component distribution on the outer rim of the nanoring are aligned (almost) perpendicularly to the $\bar{k}_p$ vector in $90^{\circ}$ ($106^{\circ}$) azimuthal orientation. In contrast to the 300 nm periodic pattern, from the two lobes only one of the tips and the larger arch is more shiny on the nanocrescents. The $E_z$ field component distribution indicates two lobes on the nanoring along / slightly rotated with respect to the $\bar{k}_p$ vector, and  asymmetrical two lobes that are more shiny on the small arch of the nanocrescents, and are even shinier than those on the 300 nm periodic pattern. The contribution of nanocrescents in quadrumer is not significant at this extremum, similarly to the 300 nm periodic pattern (Fig. \ref{IICSM_DG_cv_04}-\ref{IICSM_DG_cv_05}/a and b, bottom). In addition to this a periodic modulation also appears along the $\bar{k}_p$ vector. The characteristic charge distribution includes reversal dipolar distribution on the inner and outer rim along / slightly rotated with respect to the $\bar{k}_p$ vector, and a weak quadrupolar distribution on the nanocrescents.  The periodic charge modulation corresponds to SPP2 grating-coupled in (-2, 0) order, that is analogue to SPP1 coupled in (-1, 0) order in case of 300 nm pattern of singlet nanoring, singlet nanocrescent and same miniarray containing unit cells, but according to the two-times larger period, two periods of the charge modulation cover the unit cell. The miniarray local maximum originates again from the sum of the coincident r-U mode of the nanoring and c-U mode on the nanocrescents quadrumer, whereas SPP2 grating-coupling is responsible for the narrow global maximum (Fig. \ref{IICSM_DG_cv_06}b bottom and \ref{IICSM_DG_cv_06}c right). \\
The large / pronounced local maximum related to Wood-Rayleigh anomaly appearing at larger wavelength is exceptional, since in this case grating-coupling occurs in a spectral interval separated from the LSPR supported by components of the miniarray. Here the $B_z$ ($E_z$) field component is aligned perpendicularly (parallel) to the $\bar{E}$-field oscillation direction both in $90^{\circ}$ and $106^{\circ}$ azimuthal orientation, and in addition to this there is a strong periodic lateral modulation also on $E_z$ field component distribution. The corresponding charge distribution consists of reversal dipoles on the nanoring along the $\bar{k}_p$ vector, and dipoles on the nanocrescents parallel to that on the outer rim of the nanoring. In both orientations a lateral charge modulation is dominant with a period equal to the wavelength of SPP1 coupled in ($\pm$1, 0) order (Fig. \ref{IICSM_DG_cv_06}b, bottom and \ref{IICSM_DG_cv_06}c, right).

\subsection{Enhancement of dipolar emitters via rectangular patterns of different nano-objects}
\label{Enc_dip_emi}

\begin{figure}[h]
\center
	{\includegraphics[width=0.75\textwidth]{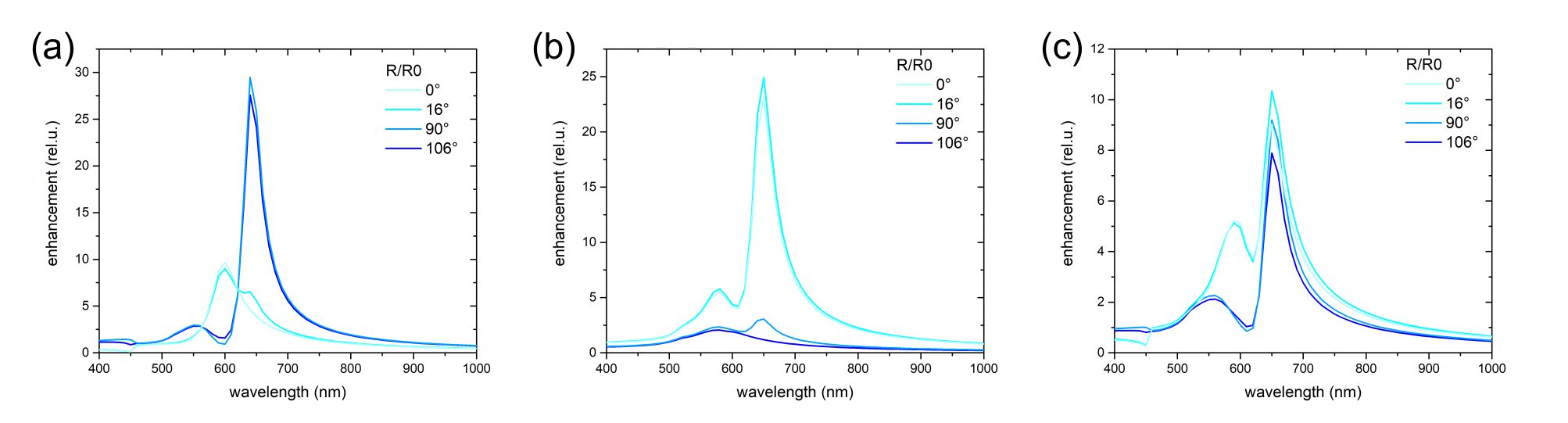}}
\caption{Radiative rate enhancement spectra of dipolar emitters achievable via 300 nm rectangular patterns consisting of (a) nanorings, (b) quadrumer of nanocrescents, (c) complex miniarray.}
\label{dip_encha_01}       
\end{figure}

When four dipoles are deepened into the conave nanorings composing a 300 nm rectangular pattern both raditaive rate enhancement spectra taken either close to / in C orientation exhibit a global maxium originating from r-C mode (600 nm), in addition to this a local maximum (640 nm) appears due to cross-coupling of a r-U mode in $16^{\circ}$ azimuthal orientation. The global maximum is forward shifted by 10 nm, whereas the local maximum is coincident with the shoulder compared to counterpart extrema on the plane wave illuminated absorptance spectra. Close to / in U orientation of quadrumer nanocrescents a local (550 nm) and a global maximum (640 nm) appears that corresponds to the r-U resonance on the nanoring and to grating-coupled SPP1, respectively. The local maximum is forward shifted by 10 nm, whereas the global maximum is coincident with the counterpart maximum on the plane wave illuminated absorptance spectra (Fig. \ref{dip_encha_01}a).\\
When four dipoles are deepened into the slightly rotated nanocrescents composing a quadrumer close to / in C orientation a local maximum appears instead of a shoulder, which is followed by a global maximum. The local maximum corresponds to the mixed c-C2 and c-C1 modes, whereas the global maximum originates from c-C1 resonance. The local (580 nm) / global (650 nm) maximum appears at a location, which is coincident / forward shifted by 10 nm with respect to the counterpart extremum on the absorptance spectra of plane wave illuminated quadrumers. Close to / in U orientation a c-U-resonance (580 nm) related local / global maximum appears, which is forward shifted by 20 nm with respect to the counterpart global maximum on the plane wave illuminated absorptance spectrum. In addition to this close to U orientation ($90^{\circ}$ azimuthal angle) a global maximum (650 nm) is observable, that originates from cross-coupled c-C1 mode and is forward shifted by 10 nm with respect to the counterpart local maximum on the plane wave illuminated spectrum (Fig. \ref{dip_encha_01}b).

In case of the rectangular pattern composed of complex miniarray the peaks on the enhancement spectra are added but almost inherit the shape of the nanoring and quadrumer spectra, which reveals that a weak interaction occurs in between the composing concave nano-objects even if they are illuminated by dipoles. Namely, local maxima (590 nm) appear at a spectral location corresponding to the r-C resonance on the nanoring overlapped with the mixed c-C2 and c-C1 resonance on the quadrumer, whereas a large global maximum appears (650 nm) which originates from the c-C1 resonance on the quadrumer and is overlapped with the cross-coupled r-U mode on the nanoring. The local maximum is coincident, whereas the global maximum is shifted by 10 nm with respect to counterpart global maxima on the absorptance spectra of plane wave illuminated miniarrays, respectively. In U orientation the local maximum (560 nm) originates from the r-U and c-U resonance on the nanoring an nanocrescent, whereas the global maximum (650 nm) originates from grating-coupled SPP1 modes. The former is coincident, whereas the latter is forward shifted by 10 nm with respect to counterpart extrema on the absorptance spectra of plane wave illuminated miniarrays (Fig. \ref{dip_encha_01}c).

\section{Discussion and conclusion}
\label{Dis_con}

The inspection of the reference hexagonal nanohole patterns uncovered the LSPRs that are at play also in rectangular nanohole patterns (Fig. \ref{IICSM_DG_cv_01}). In case of rectangular patterns the common difference between the charge and near-field distributions is the clockwise rotation in $16^{\circ}$ and $106^{\circ}$ azimuthal orientation with respect to those observable at $0^{\circ}$ and $90^{\circ}$ azimuthal angles (Fig. \ref{IICSM_DG_cv_02}-\ref{IICSM_DG_cv_06}).	\\
On the rectangular pattern of singlet concave nanorings mainly the clockwise rotation of the reversal dipoles is observable at the common extrema. In $16^{\circ}$ azimuthal orientation, which is C orientation of quadrumers, a shoulder appears caused by r-U mode cross-coupling promoted by the horizontal $\bar{E}$-field component. Surprisingly, the localized charge distribution is rotating in $16^{\circ}$ azimuthal orientation. In U orientation after the r-U mode related local maximum a global maximum appears at $90^{\circ}$/$106^{\circ}$ azimuthal angle. The dominant $\bar{E}$-field component along $\bar{k}_p$ direction results in (-1, 0) order SPP1 grating-coupling accompanied by a periodic modulation both on the $E_z$ field component and charge distribution. This periodic modulation is stronger in $90^{\circ}$ azimuthal orientation, whereas in $106^{\circ}$ azimuthal orientation both the $B_z$ and $E_z$ field component as well as the charge distribution is governed by the competition of the r-U LSPR and grating-coupled SPP1 mode (Fig. \ref{IICSM_DG_cv_02}). \\
On the rectangular pattern of horizontal singlet concave nanocrescents the c-C and c-U resonance arises in $0^{\circ}$ and $90^{\circ}$ azimuthal orientation, which results in perfect alignment of the local fields along and perpendicularly to the nanocrescents symmetry axes. In/close to C orientation at the shoulder quadrupolar and hexapolar charge distribution is also observable. However, the hexapolar modulation is less dominant in $16^{\circ}$, accordingly the $E_z$ field distribution is forward rotated with respect to the nanocrescent symmetry axis. Symmetrical / asymmetrical dipolar distribution develops in $0^{\circ}$ / $16^{\circ}$ azimuthal orientation at the global maximum. In / close to U orientation symmetrical / asymmetrical qadrupolar distribution develops both at the local and global maximum. A narrow global maximum appears that originates from the grating-coupling of SPP1 in (-1, 0) order both at $90^{\circ}$ / $106^{\circ}$ azimuthal angle. The quadrupolar LSPR is accompanied by a periodic modulation at the global maximum both on the $E_z$ field and charge distribution, which is stronger in $90^{\circ}$ azimuthal orientation due to the parallelism of the $\bar{E}$-field oscillation and $\bar{k}_p$ directions. In addition to this a shoulder appears in $106^{\circ}$ azimuthal orientation, which originates from a cross-coupled c-C1 resonance (Fig. \ref{IICSM_DG_cv_03}).\\
The C and U resonance arises in $16^{\circ}$ and $106^{\circ}$ azimuthal orientation on the rectangular pattern of quadrumers, which is capable of resulting in perfect alignment of the local fields along and perpendicularly to the symmetry axes of the slightly rotated nanocrescents. As a consequence, close to / in C orientation ($0^{\circ}$ / $16^{\circ}$ azimuthal angle) of the rectangular quadrumer pattern at the shoulder the charge distribution is asymmetrically / symmetrically quadrupolar, whereas at the global maximum it is asymmetrically / symmetrically dipolar. Close to / in U orientation ($90^{\circ}$ / $106^{\circ}$ azimuthal angle) the quadrupolar charge distribution is asymmetrical / symmetrical on each composing nanocrescents at the global maximum. The additional shoulder, which appears in $90^{\circ}$ azimuthal orientation of the quadrumer, originates from a cross-coupled c-C1 resonance (Fig. \ref{IICSM_DG_cv_04}).\\
The 300 nm periodic rectangular pattern of a concave miniarray inherits the features of the rectangular patterns composed of the singlet nanoring and the quadrumer of nanocrescents. Caused by the non-perfect alignment with respect to the $\bar{E}$-field oscillation direction the modes on the nano-objects interact close to C orientation ($0^{\circ}$ azimuthal angle) and more commensurate charge and field accumulation is observable on the nanocrescents at the local maximum and on the nanoring at the global maximum as well. In comparison, due to the perfect alignment of the fields along the symmetry axes of the nanocrecents in C orientation ($16^{\circ}$ azimuthal angle), the charge and field distribution is more dominant on the nanoring at the local maximum and on the nanocrescents at the global maximum complementary. The $E_z$ field lobes are perfectly aligned along the $\bar{E}$-field oscillation direction on the nanocrescents in C orientation both at the local and global maximum, whereas the $E_z$ field lobes are parallel (weak of less well defined directivity) to the $\bar{E}$-field oscillation direction on the nanoring at the local (global) maximum. \\
Close to U orientation ($90^{\circ}$ azimuthal angle) the sub-sets of nanocrescents are more distinguishable at the local maximum. Weak asymmetrical quadrupolar distribution develops on the nanocrescents in $90^{\circ}$ azimuthal orientation also at the global maximum, but this is accompanied by a noticeable periodic modulation both on the $E_z$ field and charge distribution due to the parallelism of the $\bar{E}$-field oscillation and $\bar{k}_p$ directions. In comparison, the sub-sets of nanocrescents are less distinguishable in U orientation ($106^{\circ}$ azimuthal angle) at the local maximum. Weak asymmetrical quadrupolar distribution develops on the nanocrescents in U orientation ($106^{\circ}$ azimuthal angle) at the global maximum, which is perturbed both on the $E_z$ field and on the charge distribution by a relatively weaker periodic modulation caused by the misalignment of the $\bar{E}$-field oscillation and $\bar{k}_p$ directions.  \\
The 600 nm periodic rectangular pattern composed of a concave miniarray exhibits similar features as the 300 nm periodic pattern. The differences between the perfect and non-perfect orientations are slightly less well defined, which can be explained by the smaller surface fraction of the nano-objects in the unit cell. In C orientation the $E_z$ field lobes are perfectly aligned along the $\bar{E}$-field oscillation direction on the nanocrescents both at the local and global maximum, whereas the $E_z$ field lobe is parallel (significantly weaker of less well defined directivity) on the nanoring at the local (global) maximum to the $\bar{E}$-field oscillation direction. In the spectral interval of the ($\pm$1, 0) order grating-coupling occurring outside the LSPR (Wood-Rayleigh anomaly) in $16^{\circ}$ azimuthal orientation the propagating modes related $E_z$ field and charge modulation is noticeable, however it is significantly stronger than the periodic modulation close to C orientation ($0^{\circ}$ azimuthal angle). \\
In U orientation the LSPRs on the nanocrescents are slightly less distinguishable at the local maximum, however at the global maximum the periodic modulation accompanying the SPP2 grating-coupled in (-2,0) order is slightly weaker both on the $E_z$ field and charge distribution. In addition to this, the grating-coupling of the SPP1 modes in ($\pm$1, 0) order through $\bar{k}_p$ lattice vector results in Wood-Rayleigh anomaly with considerable efficiency, however it is slightly smaller than that achieved close to U orientation ($90^{\circ}$ azimuthal angle).\\
The dispersion characteristics have been taken in $0^{\circ}$ and $90^{\circ}$ azimuthal orientations (Fig. \ref{IICSM_DG_cv_01}d, e and Figs. \ref{IICSM_DG_cv_02}-\ref{IICSM_DG_cv_06}/c). In present systems grating-coupling does not occur in $0^{\circ}$ azimuthal orientation in the inspected spectral interval caused by the large $\bar{k}_t$ corresponding to the small $t$ unit cell side length, except the 600 nm periodic pattern of the miniarray. In contrast the coupling in $90^{\circ}$ azimuthal orientation via $\bar{k}_p$ grating vector of the rectangular lattice results in well-defined bands in the inspected wavelength interval, except the 300 nm periodic pattern of the nanocrescent quadrumer.\\
The dispersion characteristics of the hexagonal pattern of concave nanorings does not possess azimuthal orientation dependence (Fig. \ref{IICSM_DG_cv_01}d). A well-defined and tilting independent flat band is identifiable, which corresponds to the identical r-C and r-U LSPR on the nanorings in $0^{\circ}$ and $90^{\circ}$ azimuthal orientation. In comparison, on the dispersion characteristics of the rectangular pattern of concave nanorings taken in $0^{\circ}$ azimuthal orientation a tilting independent flat band corresponding to r-C resonance is observable, whereas in $90^{\circ}$ azimuthal orientation the flat band corresponding to r-U resonance is perturbed by the SPP1 band grating-coupled in (-1, 0) order (Fig. \ref{IICSM_DG_cv_02}c).\\
The dispersion characteristics of the hexagonal pattern of concave nanocrescents, 300 nm periodic rectangular pattern of horizontal singlet concave nanocrescents and quadrumer of slightly rotated concave nanocrescents, as well as the 300 nm and 600 nm rectangular pattern of the complex miniarray exhibit well-defined and similar LSPR related azimuthal orientation dependence (Fig \ref{IICSM_DG_cv_01}e, Fig \ref{IICSM_DG_cv_03}-\ref{IICSM_DG_cv_06}/c). \\
In $0^{\circ}$ azimuthal orientation of the hexagonal pattern as well as of the rectangular pattern of singlet horizontal nanocrescents and quadrumer of slightly rotated nanocrescents a tilting independent strong flat band indicates the c-C1 LSPR, whereas the interaction of c-C2 and c-C1 modes results in a ghost flat band at slightly smaller wavelength. This interaction related band is less / more well defined in case of a horizontal singlet / slightly rotated quadrumer nanocrescent in a rectangular pattern (Fig. \ref{IICSM_DG_cv_01}e, Figs \ref{IICSM_DG_cv_03}-\ref{IICSM_DG_cv_04}/c, left). In $90^{\circ}$ azimuthal orientation of the rectangular patterns the flat band corresponding to c-U resonance is perturbed by the band of SPP1 grating-coupled in (-1,0) order in presence of singlet nanocrescents, whereas there is no coupled SPP band in presence of quadrumer nanocrescents (Fig. \ref{IICSM_DG_cv_01}e, Figs \ref{IICSM_DG_cv_03}-\ref{IICSM_DG_cv_04}/c, right). \\
The dispersion characteristics of the 300 nm and 600 nm periodic rectangular patterns of miniarrays are more complex. Similarly to the composing nano-objects, in $0^{\circ}$ azimuthal orientation of the 300 nm and 600 nm periodic rectangular patterns of their miniarray a tilting independent weak flat band indicates the c-C1 LSPR on the nanocrescents, whereas the interaction of c-C2 and c-C1 modes results in a flat band, which overlaps with the tilting and azimuthal orientation independent strong band of the r-C LSPR on the nanoring (Fig. \ref{IICSM_DG_cv_05}-\ref{IICSM_DG_cv_06}/c, left). In addition to this, on the 600 nm periodic rectangular pattern the Rayleigh anomaly related band is noticeable due to the ($\pm$1, 0) order coupling of SPP1 along the $\bar{k}_p$ vector of the periodic pattern, which results in a tiny peak also at perpendicular incidence in $16^{\circ}$ azimuthal orientation  (Fig \ref{IICSM_DG_cv_05}-\ref{IICSM_DG_cv_06}/c, left).\\
In $90^{\circ}$ azimuthal orientation of the 300 nm and 600 nm periodic pattern of the complex miniarrays the tilting independent strong band corresponding to the coincident r-U LSPR on the nanoring and c-U LSPR on the nanocrescents is perturbed by SPPs coupled in (-1, 0) and (-2, 0) order, respectively. In addition to this, on the 600 nm periodic rectangular pattern the Wood-Rayleigh anomaly related SPP bands also appears due to the efficient ($\pm$1, 0) order coupling of plasmonic modes along the $\bar{k}_p$ vector of the periodic pattern, which result in significant / considerable peak at perpendicular incidence in $90^{\circ}$ / $106^{\circ}$ azimuthal orientation  (Fig \ref{IICSM_DG_cv_05}-\ref{IICSM_DG_cv_06}/c, right).\\
The peaks corresponding to SPP coupling can be distinguished from LSPRs based on two important differences: (i) The periodic charge distribution (as well as the related $E_z$ field component distribution) is parallel to the $\bar{k}_p$ vector rather than is governed by the $\bar{E}$-field oscillation direction, since the latter determines only the charge separation of LSPR. (ii) The period of the modulation appearing both on the $E_z$ field component and charge distribution equals to the wavelength of the SPP mode grating-coupled in (-1, 0) and (-2, 0) order in $90^{\circ}$ azimuthal orientation of the 300 nm and 600 nm periodic pattern in the spectral interval overlapping with LSPR (640 nm and 630 nm), or (weakly) strongly grating-coupled in ($\pm$1, 0) order in ($16^{\circ}$)   $90^{\circ}$ and $106^{\circ}$azimuthal orientation of the 600 nm periodic pattern in absence of LSPR (970 nm).\\
Our present study proves that significant enhancement of dipolar emitters is achievable in spectral intervals of plasmonic resonances on complex concave patterns that are tuneable by the integrated lithography.

\begin{acknowledgements}
This work was supported by the National Research, Development and Innovation Office (NKFIH) “Optimized nanoplasmonics” (K116362) and European Union, co-financed by the European Social Fund. “Ultrafast physical processes in atoms, molecules, nanostructures and biological systems” (EFOP-3.6.2-16-2017-00005). Á. Sipos gratefully acknowledges the support of NKFIH PD-121170.
\end{acknowledgements}

%
%



\end{document}